\documentclass[aip,jmp, amsmath,amssymb, preprint]{revtex4-1}

\usepackage{graphicx}% Include figure files
\usepackage{dcolumn}% Align table columns on decimal point
\usepackage{bm}% bold math
\usepackage{mathtools}
\usepackage{amsmath}
\usepackage{amssymb}
\usepackage{amsthm}
\usepackage{mathrsfs}
\usepackage{yfonts}
\usepackage{txfonts}

\newcommand{\M}[0]{\ensuremath{\mathbf{M}}}
\newcommand{\T}[0]{\ensuremath{\mathrm{T}}}
\newcommand{\U}[0]{\ensuremath{\mathrm{U}}}
\newcommand{\SU}[0]{\ensuremath{\mathrm{SU}}}

\newcommand{\g}[0]{\ensuremath{\mathbf{g}}}
\newcommand{\q}[0]{\ensuremath{\mathbf{q}}}
\newcommand{\uu}[0]{\ensuremath{\mathbf{u}}}
\newcommand{\vv}[0]{\ensuremath{\mathbf{v}}}

\newcommand{\R}[0]{\ensuremath{\mathbb{R}}}

\newcommand{\D}[0]{\ensuremath{\mathcal{D}}}
\newcommand{\HD}[0]{\ensuremath{\hat{\mathcal{D}}}}
\newcommand{\dd}[0]{\ensuremath{\mathrm{d}}}
\newcommand{\hd}[0]{\ensuremath{\hat{\mathrm{d}}}}

\newcommand{\ee}[0]{\ensuremath{\mathbf{e}}}
\newcommand{\EE}[0]{\ensuremath{\mathbf{E}}}
\newcommand{\AAA}[0]{\ensuremath{\mathbf{A}}}
\newcommand{\HAA}[0]{\ensuremath{\hat{\mathbf{A}}}}
\newcommand{\LL}[0]{\ensuremath{\mathbf{L}}}
\newcommand{\Lg}[0]{\ensuremath{\mathscr{L}}}
\newcommand{\Mkw}[0]{\ensuremath{\mathscr{M}}}

\newcommand{\RR}[0]{\ensuremath{\mathbf{R}}}
\newcommand{\HR}[0]{\ensuremath{\hat{\mathbf{R}}}}
\newcommand{\TT}[0]{\ensuremath{\mathbf{T}}}
\newcommand{\KK}[0]{\ensuremath{\mathbf{K}}}

\newcommand{\SO}[0]{\ensuremath{\mathbf{SO}}}

\newcommand{\GL}[0]{\ensuremath{\mathbf{GL}}}
\newcommand{\OOO}[0]{\ensuremath{\mathbf{O}}}
\newcommand{\Td}[0]{\ensuremath{\mathbb{T}}}
\newcommand{\Ad}[0]{\ensuremath{\mathbb{A}}}

\newcommand{\hs}[0]{\hspace*{5.9mm}}

\newcommand{\bleps}[0]{\ensuremath{\boldsymbol{\varepsilon}}}
\newcommand{\blome}[0]{\ensuremath{\boldsymbol{\omega}}}
\newcommand{\BSigma}[0]{\ensuremath{\boldsymbol{\Sigma}}}
\newcommand{\bsigma}[0]{\ensuremath{\boldsymbol{\sigma}}}
\newcommand{\blpi}[0]{\ensuremath{\boldsymbol{\pi}}}
\newcommand{\blPI}[0]{\ensuremath{\boldsymbol{\Pi}}}
\newcommand{\blphi}[0]{\ensuremath{\boldsymbol{\varphi}}}
\newcommand{\blPHI}[0]{\ensuremath{\boldsymbol{\Phi}}}
\newcommand{\UU}[0]{\ensuremath{\mathbf{U}}}
\newcommand{\pp}[0]{\ensuremath{\mathbf{p}}}
\newcommand{\PP}[0]{\ensuremath{\mathbf{P}}}
\newcommand{\bb}[0]{\ensuremath{\mathbf{b}}}
\newcommand{\BB}[0]{\ensuremath{\mathbf{B}}}
\newcommand{\CC}[0]{\ensuremath{\mathbf{C}}}
\newcommand{\YY}[0]{\ensuremath{\mathbf{Y}}}
\newcommand{\Y}[0]{\ensuremath{\mathcal{Y}}}
\newcommand{\XX}[0]{\ensuremath{\mathbf{X}}}
\newcommand{\X}[0]{\ensuremath{\mathcal{X}}}
\newcommand{\xxx}[0]{\ensuremath{\mathbf{x}}}
\newcommand{\y}[0]{\ensuremath{\mathbf{y}}}

\newcommand{\HBB}[0]{\ensuremath{\hat{\mathbf{B}}}}
\newcommand{\HCC}[0]{\ensuremath{\hat{\mathbf{C}}}}
\newcommand{\HH}[0]{\ensuremath{\mathbf{H}}}
\newcommand{\FFF}[0]{\ensuremath{\mathbf{F}}}
\newcommand{\GGamma}[0]{\ensuremath{\hat{\boldsymbol{\Gamma}}}}

\newcommand{\vb}[1]{\ensuremath{\big|_{#1}}}
\newcommand{\A}[0]{\ensuremath{\mathcal{A}}}
\newcommand{\B}[0]{\ensuremath{\mathcal{B}}}

\newcommand{\C}[0]{\ensuremath{\mathcal{C}}}
\newcommand{\LLC}[0]{\ensuremath{\LL^{(\text{EC})}}}
\newcommand{\LLR}[0]{\ensuremath{\LL^{(\text{Rest})}}}
\newcommand{\HHC}[0]{\ensuremath{\HH^{(\text{EC})}}}
\newcommand{\HHR}[0]{\ensuremath{\HH^{(\text{Rest})}}}
\newcommand{\ppkol}{\pp^{^{\bot}}}
\newcommand{\pprov}{\pp^{^{||}}}
\newcommand{\pkol}[1]{\tilde{p}^{^{\bot}#1}}
\newcommand{\prov}[2]{\tilde{p}^{^{||}#1}_{\phantom{^{||}}\!#2}}
\newcommand{\PPkol}{\PP^{^{\bot}}}
\newcommand{\PProv}{\PP^{^{||}}}
\newcommand{\Pkol}[2]{\tilde{P}^{^{\bot}#1}_{\phantom{^{\bot}}\!#2}}
\newcommand{\Prov}[1]{\tilde{P}^{^{||}}_{\phantom{^{||}\!}#1}}
\newcommand{\tsigma}{\tilde{\sigma}}
\newcommand{\ints}{\int\limits_{\Sigma}}

\newcommand{\ham}{\mathsf{H}^{\text{(EC)}}}

\newcommand{\Conf}{\mathfrak{Conf}}

\begin{document}
%\preprint{AIP/123-QED}

\title[]{On Einstein-Cartan Theory: I. Kinematical description} %Title of paper


\author{Mari\' an Pilc}
\email[]{marian.pilc@gmail.com}
\noaffiliation
\affiliation{Institute of Theoretical Physics,Faculty of Mathematics and Physics, Charles University,V Holesovickach 2,
180 00 Praha 8, Czech Republic}
%\thanks{}
%\noaffiliation is required (may also be used with the \author command).

\date{\today}

\begin{abstract}
Equations of motion for general gravitational connection and orthonormal coframe from the Einstein-Hilbert
type action are derived. Our formulation does not fix coframe to be tangential to spatial section hence Lorentz group is still present as a part of gauge freedom. 3+1 decomposition introduces tangent Minkowski structures hence Hamilton-Dirac approach to dynamics works with Lorentz connection over the spatial section. The second class constraints are analyzed and Dirac bracket is defined. Reduction of
the phase space is performed and canonical coordinates are introduced.
\end{abstract}

%\pacs{123456}% insert suggested PACS numbers in braces on next line
\maketitle %\maketitle must follow title, authors, abstract and \pacs

%----------------------------------------------------
%						Introduction
%----------------------------------------------------

\section{Introduction}\label{Section-Introduction}

Einstein theory of General Relativity is well known and understood theory of gravitation
for almost a century. Gravitational interaction is described by metric tensor $\g$ of a spacetime manifold 
$\M$. Einstein needed to assume that gravitational connection is metric-compatible and torsion-free in his
derivation of equations of motion for $\g$. We will call such connection geometrical or Riemann-Levi-Civita 
(RLC). Important thing is that the geometrical connection $^{(RLC)}\hat{\nabla}$ is uniquely determined by
the metric $\g$. If we want to describe gravitational system interacting with Dirac field within General 
Relativity then we should express metric in terms of orthonormal coframe $\ee^a$. The action of such system 
is given by sum of Einstein-Hilbert and Dirac actions\cite{RLC-electron}, where the spacetime external 
derivative operator $\hd$ should be replaced by $^{(RLC)}\hat{\nabla}$ in order to have a final theory 
locally Lorentz invariant.\\ 
\hs There exist another approach to the relativistic theory of gravitation. Interaction in Standard model 
of elementary particles is described by gauge potentials given by the appropriate gauge group, e.g. $\U(1)$ for 
electromagnetism or $\SU(2)$ for electroweak theory, etc. Similar ideas as in Standard model can be used in 
construction of gravitational theory. Kibble\cite{Kibble} used Poincar\' e symmetry of Minkowski spacetime and 
he obtained the theory where gravitational potential is given by general metric-compatible connection. In contrast
to General Theory of Relativity the condition of vanishing torsion is given by equation of motion and the rest
of dynamics is described by Einstein equation. In other words in the case of pure gravity these two theories 
are physically equivalent. Since spacetime is no longer flat the Poincar\' e group is no longer global symmetry of
solution hence only local Poincar\' e symmetry plays a role of gauge group of Kibble theory\cite{Hehl1}. 
There exists another generalization\cite{Hehl2} of this approach called theory of affine connection, where 
Lorentz group is replaced by $\GL(4)$.\\
\hs We will see in this article, that these three theories are physically equivalent at least in the case of pure 
gravity. Problems occur if we want to add matter fields with Lagrangian depending on connection 1-forms. 
In general case these three theories are no longer equivalent. We will show that general connection can be 
decomposed into metric-compatible connection plus something. If matter Lagrangian depends only on 
metric-compatible part of general connection then Kibble theory and theory of affine connection are 
equivalent as we will see in last paper of this series. Physically reasonable example of such matter are 
all Standard model fields. Bosons do not interact directly with connection while spinor part of Lagrangian 
depends only on metric-compatible part of connection. Since Standard model plus Gravity is everything what 
we know about the Nature at the present stage of Physics we will not distinguish between these two theories 
and we will call them as Einstein-Cartan theory in this series except the next section and the last paper 
for simplicity, where we will show an example of Bi\v c\' ak vector field\cite{Bicak_pole}, which violates 
this equivalence and one must consider three different descriptions of its interaction with gravity.\\
\hs We will focus on Einstein-Cartan theory in this paper. The motivation for this choice 
is taken from loop quantum gravity, where Ashtekar \nobreak{connection} $\AAA$ on a spatial section $\Sigma$ 
is defined by RLC connection of $\q$ ($\q$ is a metric on $\Sigma$ induced from the 4-dimensional metric $\g$ 
of the spacetime $\M$) and an external curvature of the 4-dimensional RLC connection. Ashtekar originally 
began with complex connection $\AAA$ but problems with reality conditions or hermiticity of inner product of 
quantum Hilbert space caused that Barbero-Immirzi parameter enters the theory and $\AAA$ becomes real. This 
parameter plays no role on classical level, but after quantization it causes ambiguity and must be fixed by 
comparison of Hawking-Bekenstein entropy with entropy computed from loop theory. Fermionic matter was 
successfully added to loop gravity only on kinematical level and problem of dynamics remains unresolved. 
And last but not least, problem is that general theory is $\SO(\g)$ invariant what is still true in 
the case of complex Ashtekar connection but the real loop theory broke down this explicit invariance 
to $\SO(\q)$ \cite{Samuel-kritika1}$^,$\cite{Samuel-kritika2}.\\ 
\hs If one\cite{Livine} does not fix coframe to be tangential to $\Sigma$ in opposite to euclidean loop 
gravity then all degrees of freedom enters the theory which can then be expressed 
as $\SO(\g)$ gauge theory. As is shown in \ref{2+1GR} this leads to the theory where torsion 
appears as the first class constraint in the case of 2+1 dimensional gravity what is good news for 2+1 
dimensional theoretical physicists, because they can work with $\SO(\g)$ gauge connection instead of 
2+1 analogue of Ashtekar connection and problem of vanishing torsion can be solved on quantum level as they 
wish. Unfortunately in the case of 3+1 dimensional  Einstein-Cartan theory the condition of vanishing 
the torsion is split in two parts where one is the first class and other is the second class constraints. 
Therefore new potential problems like introduction of ghosts might be solved on quantum level.\\
\hs  This is the first part of a series of three papers devoted to  Einstein-Cartan theory.
In this paper, we will focus on the derivation of Hamiltonian-Dirac formulation of our physical system. The
paper is organized as follows. In section \ref{Section-Lagrangian_of_Cartan_Theory}, Lagrangian 
formulation of  the Einstein-Cartan theory is formulated in the language of forms valued in 
the tangent tensor algebra $\Lambda\Td\M$. Equations of motion (EOM) are derived and equivalence between 
theory of General Relativy and  Einstein-Cartan theory is also shown in this section. 3+1 decomposition is 
performed in section \ref{Section-3+1_Decomposition} and also some useful formulas are \nobreak{evaluated} 
there. In section \ref{section-Hamiltonian}, the Hamiltonian of the theory is written and 
separation of constraints into the first and second class is performed. In section \ref{Section-Dirac_brackets}, 
Dirac brackets are introduced and coordinates on the reduced phase space are defined. And in the last section 
\ref{Section-Discussion_and_Open_Problems}, open problems are discussed and possible solutions are sugessted. 
Also few comments are added about possible quantization.

%----------------------------------------------------------------------------------------------------------
%------------------------- Lagrangian of Cartan Theory-----------------------------------------------------
%----------------------------------------------------------------------------------------------------------
\section{Lagrangian of  Einstein-Cartan Theory}\label{Section-Lagrangian_of_Cartan_Theory}
Let $(\M=\R[t]\times\Sigma,\g)$ be a spacetime manifold equipped with metric $\g$ (signature$(\g)=(+,-,-,-)$). 
Geroch's theorem \cite{Geroch} says that a spinor structure over the manifold $\M$ exists iff there 
exists a global 
orthonormal frame $\ee_a$ over $\M$ and $\M$ is orientable. These two conditions restrict possible 
topological shapes of $\M$ and $\Sigma$, e.g. if the spacetime manifold is given by product 
$\M=\R\times$"3-dimensional sphere" then Geroch's conditions are not fulfilled and the spinor structure 
can not be defined over such manifold, in other words if one considers Friedman's models then 
closed model violates Geroch's conditions. We assume Geroch's conditions already now in the case of 
pure gravity since spinors should be added into the theory later so there is no loss of generality
\footnote{One may say that we can define the spinor structure locally and work with such structure.
But there  may occur some certain phatological features. We will not focus our attention to this problem. 
Therefore {"}no loss of generality{"}.}.\\ 
\hs Let  $\ee_a$ be a global orthonormal frame and $\ee^a$ its dual. Then every useful geometrical or 
\nobreak{gravitational} variable can be written in a global manner. Let us look at the basic quantities:\\
metric ($\eta_{ab}$ is Minkowski matrix: $(\eta_{ab})=\text{diag}(+1,-1,-1,-1)$)
\begin{eqnarray}
\g=\eta_{ab}\ee^a\otimes\ee^b,\nonumber
\end{eqnarray}
4-volume form ($\bleps_{abcd}$ is Levi-Civita antisymmetric symbol, see convention in \nobreak{appendix} A)
\begin{eqnarray}
\hat{\BSigma}=\frac{1}{4!}\bleps_{abcd}\ee^a\wedge\ee^b\wedge\ee^c\wedge\ee^d,\nonumber
\end{eqnarray}
gravitational connection 1-form $\GGamma^b_{\phantom{b}a}$ ($\uu$ is arbitrary vector)
\begin{eqnarray}
\hat{\nabla}_{\uu}\ee_a=\GGamma^b_{\phantom{b}a}(\uu)\ee_b\nonumber
\end{eqnarray}
or its curvature 2-form
\begin{eqnarray}
\hat{\FFF}^a_{\phantom{b}b}=\hat{\dd}\GGamma^a_{\phantom{b}b}+\GGamma^a_{\phantom{b}c}\wedge\GGamma^c_{\phantom{b}b}.\nonumber
\end{eqnarray}
General Relativity sets connection $\hat{\nabla}$ to be geometrical and the Einstein-Hilbert action of GR is
\begin{eqnarray}
S_{\text{EH}}=\int-\frac{1}{16\pi\kappa}R_{\g}\omega_{\g}\nonumber
\end{eqnarray}
where $R_{\g}$ is a Ricci scalar related to the RLC connection of metric tensor $\g$, 
$\omega_{\g}=\sqrt{-\det|g|}\dd^4x$ is its volume form and $\kappa$ is Newton's constant (c=1). The action 
written in this form explicitly 
depends on the choice of coordinates and one should overlap few coordinate's neighbourhoods and 
solve boundary terms if one wants to cover the whole manifold $\M$ in general case. But if we use our 
assumption about $\ee^a$ then we can rewrite the Einstein-Hilbert action into the following 
geometrical form
\begin{eqnarray}
S_{\text{EH}}=\int-\frac{1}{32\pi\kappa}\bleps_{abcd}\eta^{b\bar{b}} \RR^{\phantom{\g}a}_{\g\phantom{b}\bar{b}}\wedge\ee^c\wedge\ee^d,\label{EH-action}
\end{eqnarray}
where $\RR^{\phantom{\g}a}_{\g\phantom{a}b}$ is a curvature 2-form of RLC connection.
The action (\ref{EH-action}) is a functional of basic variables $\ee^a=e^a_{\mu}\dd x^{\mu}$
($\mu=0,\, \alpha=0,1,2,3$ are spacetime coordinate's indices) and we should make variation of 
the action with respect to them. The idea of the theory of general linear connection
is very simple, gravitational connection $\hat{\nabla}$ is no more geometrical. 
Variation should be made independently in both variables $\ee^a$ and $\GGamma^{a}_{\phantom{a}b}$ in action 
being of Einstein-Hilbert type 
\begin{eqnarray}
S=\int\limits_{\Omega}-\frac{1}{32\pi\kappa}\boldsymbol{\varepsilon}_{abcd}
\eta^{b\bar{b}}\hat{\FFF}^{a}_{\phantom{a}\bar{b}}\wedge\ee^c\wedge\ee^d,\label{Cartan_action}
\end{eqnarray}
where $\Omega$ is a timelike compact set, i.e. $\Omega=<t_{i};t_f>\times\Sigma$. For simplicity we assume in this paper 
that $\Sigma$ is compact manifold, e.g. torus; our next paper will be focused also on noncompact manifolds 
with boundary. Let us decompose variable $\GGamma^a_{\phantom{a}b}$ into $\OOO(\g)$-irreducible parts
\begin{eqnarray}
\GGamma^{ab}=\eta^{bc}\GGamma^{a}_{\phantom{a}c}=\HAA^{ab}+\HBB\eta^{ab}+\HCC^{ab}\label{rozklad_premenne}
\end{eqnarray}
where $\HAA^{ab}$ is antisymmetric and $\HCC^{ab}$ is symmetric and traceless 1-form, respectively. The 
curvature $\hat{\FFF}^a_{\phantom{a}b}$ can be expressed as
\begin{eqnarray}
\hat{\FFF}^{ab}=\eta^{bc}\hat{\FFF}^{a}_{\phantom{a}c}=\HR^{ab}+\hat{\dd}\HBB\eta^{ab}+\HD\HCC^{ab}+\eta_{cd}\HCC^{ac}\wedge\HCC^{db}\nonumber
\end{eqnarray}
where $\HD$ is a metric-compatible connection defined on $\Lambda\Td\M$ (see notation in appendix A) by 
$\HD\uu^a=\dd \uu^a+\eta_{bc}\HAA^{ab}\wedge\uu^c$ for $\forall \uu^a\in\Lambda\Td^1\M$ and $\HR^{ab}$ 
is its curvature.
If $\tilde{\ee}^a=O^a_{\phantom{a}\bar{a}}\ee^{\bar{a}}$ is a new coframe with $O^a_{\phantom{a}b}$ being 
Lorentz transformation, then $\HAA^{ab}$ transforms as
\begin{eqnarray}
\tilde{\HAA}^{ab}=O^a_{\phantom{a}{\bar{a}}}O^b_{\phantom{b}{\bar{b}}}\HAA^{\bar{a}\bar{b}}+O^a_{\phantom{a}{\bar{a}}}\eta^{\bar{a}\bar{b}}\dd O^b_{\phantom{b}{\bar{b}}}\nonumber 
\end{eqnarray}
while $\HBB$ and $\HCC^{ab}$ transform like tensors in their indices. The action (\ref{Cartan_action})
can be written in new variables $(\ee^a, \HAA^{ab}, \HBB, \HCC^{ab})$ as
\begin{eqnarray}
S=\int\limits_{\M}-\frac{1}{32\pi\kappa}\boldsymbol{\varepsilon}_{abcd}\HR^{ab}\wedge\ee^c\wedge\ee^d+
\int\limits_{\M}-\frac{1}{32\pi\kappa}\eta_{\bar{a}\bar{b}}\boldsymbol{\varepsilon}_{abcd}\HCC^{a\bar{a}}
\wedge\HCC^{\bar{b}b}\wedge\ee^c\wedge\ee^d\label{Cartan_action_2}
\end{eqnarray}
Notice that variable $\HBB$ does not enter the action (\ref{Cartan_action_2}). Thus variation of 
(\ref{Cartan_action_2}) with respect to $\HBB$ vanishes identically and no 
corresponding equation of motion arises, i.e.
\begin{eqnarray}
\delta_{\HBB}S=0,
\end{eqnarray}
hence $\HBB$ is strictly gauge variable. Now if we make variation of (\ref{Cartan_action_2}) with respect 
to $\HCC^{ab}$ then we get
\begin{eqnarray}
\delta_{\HCC}S=\int-\frac{1}{16\pi\kappa}\eta_{\bar{a}\bar{b}}\bleps_{abcd}\delta\HCC^{a\bar{a}}
\wedge\HCC^{b\bar{b}}\wedge\ee^c\wedge\ee^d=0 \label{hcc}
\end{eqnarray}
for $\forall\delta\HCC^{ab}$: $\delta\HCC^{ab}=\delta\HCC^{ba}$ and $\eta_{ab}\delta\HCC^{ab}=0$. Equation (\ref{hcc}) is 
equivalent to
\begin{eqnarray}
\HCC^{ab}=0\label{hcc2}.
\end{eqnarray}
If we use this fact then the action (\ref{Cartan_action_2}) can be reduced as
\begin{eqnarray}
S'=\int\limits_{\Omega}-\frac{1}{32\pi\kappa}\boldsymbol{\varepsilon}_{abcd}\HR^{ab}
\wedge\ee^c\wedge\ee^d,\label{Cartan_action_metric}
\end{eqnarray}
what is an action of Einstein-Cartan theory for metric-compatible connection $\HAA^{ab}$,
but our configuration space is little bit bigger since it is described by variables $\ee^a$, 
$\HAA^{ab}$, $\HBB$ ($\HCC^{ab}=0$ by equation (\ref{hcc}) or (\ref{hcc2})) and their velocities. Hence
we get Einstein-Cartan theory by gauge fixation $\HBB=0$. Variation of the action (\ref{Cartan_action_metric})
(see, e.g., [\cite{Fecko}]) is
\begin{eqnarray}
\delta_{\HAA,\ee} S'=\int\limits_{\Omega }\left( \frac{1}{16\pi\kappa}\bleps_{abcd}\delta\HAA^{ab}\wedge\ee^c\wedge\HD\ee^d-\frac{1}{16\pi\kappa}\bleps_{abcd}\delta\ee^a\wedge\HR^{bc}\wedge\ee^d\right)\nonumber
\end{eqnarray}
and equations of motion are
\begin{eqnarray}
0&=&\,\,\,\,\,\frac{1}{8\pi\kappa}\bleps_{abcd}\ee^c\wedge\HD\ee^d\,\,=
-\frac{1}{8\pi\kappa}\left(\hat{T}^c_{ab}+\hat{T}^d_{da}\delta^c_b-\hat{T}^d_{db}\delta^c_a\right)\hat{\BSigma}_c,\label{TorEQ}\\
0&=&-\frac{1}{16\pi\kappa}\bleps_{abcd}\HR^{bc}\wedge\ee^d=-\frac{1}{8\pi\kappa}\hat{G}^c_{\phantom{c}a}\hat{\BSigma}_c,\label{EEQ}
\end{eqnarray}
where the torsion components are given by
\begin{eqnarray}
\HD\ee^a=\hat{\TT}^a=\frac{1}{2}\hat{T}^a_{bc}\ee^b\wedge\ee^c,\nonumber
\end{eqnarray}
3-volume forms
\begin{eqnarray}
\hat{\BSigma}_a=\frac{1}{3!}\bleps_{abcd}\ee^{b}\wedge\ee^{c}\wedge\ee^d\nonumber,
\end{eqnarray}
and $\hat{G}^a_{\phantom{a}b}$ is Einstein tensor
\begin{eqnarray}
\hat{G}^a_{\phantom{a}b}&=&\hat{R}^{ca}_{\phantom{ab}cb}-\frac{1}{2}\hat{R}^{cd}_{\phantom{cd}cd}\delta^a_b,\nonumber\\
\HR^{ab}&=&\frac{1}{2}\hat{R}^{ab}_{\phantom{ab}cd}\ee^c\wedge\ee^d.\nonumber
\end{eqnarray}
Equation (\ref{TorEQ}) implies that connection $\HD$ is torsion-free and together with 
metricity of $\HD$ we have that $\HD$ is geometrical connection. Equations (\ref{EEQ}) are Einstein equations 
of General Relativity. Solution for general gravitational connection $\GGamma^{ab}$ is
\begin{eqnarray}
\GGamma^{ab}=\HAA^{ab}+\HBB\eta^{ab}\label{solution_shape},
\end{eqnarray}
where $\HBB$ is arbitrary 1-form and $\HAA^{ab}$, $\ee^a$ are given by equations (\ref{TorEQ}) 
and (\ref{EEQ}). Connection of type (\ref{solution_shape}) will be called Cartan connection in this series.
Ambiguity of $\GGamma^{ab}$ due to $\HBB$ represents an additional gauge freedom in $\GGamma^{ab}$. 
The spacetime is given by topology of $\Sigma$ which is established 
initially and the metric $\g=\eta_{ab}\ee^a\otimes\ee^b$ is given just by knowledge of $\ee^a$, 
hence $\HBB$ does not affect geometry. Thus General Relativity, Einstein-Cartan theory and theory of general linear connection are physically equivalent, at least in the case of pure gravity. We will see in last
paper of this series that equivalence of these three theories are no longer valid if we add matter Langrangian
depending on $\Gamma^a_{\phantom{a}b}$. On the other hand if matter Lagrangian depends only on metric connection
$\HAA^{ab}$, e.g. all Standard model matter especially spinor fields,  then Einstein-Cartan theory and the theory of general linear connection are physically equivalent. Since this series is focused especially on pure gravitational system or gravitational system interacting with spinor fields we will not distinguish between them and we will call both of them Einstein-Cartan theory (ECT).

%---------------------------------------------------------------------
%------------------------- 3+1 Decomposition -------------------------
%---------------------------------------------------------------------

\section{3+1 Decomposition}\label{Section-3+1_Decomposition}

We already assumed that the spacetime $\M$ is given by product $\R\times\Sigma$. This assumption is 
equivalent to the existence of a global Cauchy surface and hence solution of equations (\ref{TorEQ}) 
and (\ref{EEQ}) can be evolved from initial data on $\Sigma$ uniquely upto gauge transformation\footnote{One 
equation is still missing as we will see at the end of this section. But this equation is conservation of 
constraints given by (\ref{TorEQ}) and (\ref{EEQ}).}. Our basic variables $\ee^a$, $\HBB$ 
and $\HCC^{ab}$ belong to the algebra $\Lambda\Td\M$ while $\HAA^{ab}$ are connection forms on $\M$, so it 
will be useful to preserve this structure even in Hamiltonian formulation. Since we assume that Geroch's 
conditions are valid, there exists global orthonormal frame $\ee_a$. Let $x\in\Sigma$ then 
$\Mkw_x=\text{Span}\{\ee_a\vb{x}\}$ together with metric $\g\vb{x}$ define tangent Minkowski space at point $x$. 
Since $x$ is arbitrary point of $\Sigma$ then space $\Mkw=\cup_{x\in\Sigma}\Mkw_x$ plays analogue role 
as $\Td^1\Sigma$ but it is little bit bigger since $\Mkw$ contains even non tangential vectors. Important 
thing is that $\Mkw$ can be represented as $\hat{\Mkw}=F(\Sigma)^4$ and it is also equipped with 
Minkowski metric $\eta_{ab}$. Hat over $\Mkw$ will be omitted from now and space $\Mkw$ and its representation 
will be identified. $\Mkw$ is a vector space and we can define its tensor algebra $\Td\Mkw$ and algebra of forms 
on $\Sigma$ valued in this space $\Lambda\Td\Mkw$. Let $\ee_a$ and $\tilde{\ee}_a$ 
be two orthonormal frames in $\Mkw$. Then due to Geroch's conditions there exists just one 
$g\in\OOO(\g)\times\Sigma$ such that $\tilde{\ee}_a=g^*\ee_a$. Thus we see that there exists trivial
 principal bundle $\OOO(\g)\Sigma=\Sigma\times\OOO(\g)$ over $\Sigma$ which can be identified with $\Mkw$. 
Now we can start detail analysis of 3+1 decomposition of our variables.\\
\hs Let $\hat{\TT}\in\Lambda\Td\M$ be a $p$-form valued in $\Td\M$, then $\hat{\TT}$ can be uniquelly 
decomposed into pure spatial $(p-1)$-form $\check{\TT}$ and $p$-form $\TT$ valued in $\Mkw$
\begin{eqnarray}
\hat{\TT}=\check{\TT}\wedge\dd t+{\TT}.\nonumber
\end{eqnarray}
Another important geometric object is an external derivative operator. Let us denote by $\hat{\dd}$ external 
derivative on $\M$ while we keep $\dd$ for $\Sigma$. Anyway we still write $\dd t$ with the hope that this 
will not cause any problem. Let us apply $\hat{\dd}$ on $\hat{\TT}$, we obtain
\begin{eqnarray}
\hat{\dd}\hat{\TT}=\dd\check{\TT}\wedge\dd t+\dd t\wedge\dot{\TT}+\dd\TT,\nonumber
\end{eqnarray}
where dot means action of Lie derivate along $\partial_t$ which is just simple time derivative of components, 
e.g. for spatial 1-form $\dot{\TT}=\partial_t\T_{\alpha}\dd x^{\alpha}$, etc.
So we can project spacetime p-form onto the pure spatial p-form and (p-1)-form on $\Sigma$ and even
$3+1$ dimensional  external derivative is also writen in the language of spatial forms and their 
time and spatial derivatives.\\
\hs Let us explore what happens with orthonormal coframe $\ee^a$. We can write
\begin{eqnarray}
\ee^a=\lambda^a\dd t+\EE^a=\lambda^a\dd t+ E^{a}_{\alpha}\dd x^{\alpha},\label{ee^a}
\end{eqnarray}
where $\alpha,\beta,\gamma,\,.\,.\,.=1,2,3$  are spatial coordinate indices while $a,b,c,\,.\,.\,.=0,1,2,3$ 
are reserved for tensors on $\Mkw$.  
It is useful for our purposes to decompose even frame $\ee_a$ into spatial and time parts
\begin{eqnarray}
\ee_a=\lambda_a\partial_t+\EE_a=\lambda_a\partial_t+E_a^{\alpha}\partial_{\alpha}.\label{ee_a}
\end{eqnarray}
It should be noted that $\lambda_a\neq\eta_{ab}\lambda^a$. We hope that this notation is not 
confusing since if we need to in$/$de-crease indices then it will be explicitly written 
using metric tensor. We have $\ee^a(\ee_b)=\delta^a_b$ what is
\begin{eqnarray}
\begin{matrix}
		\begin{pmatrix}
		\lambda^a & E^a_{\alpha}
		\end{pmatrix}\\
		\begin{matrix}
		\end{matrix}
\end{matrix}
\begin{pmatrix}
	 \lambda_b\\
	 E_b^{\alpha}
\end{pmatrix}
=\lambda^a\lambda_b+E^a_{\alpha}E^{\alpha}_b=\delta^a_b,\label{pr1}
\end{eqnarray}
thus matrices $(\lambda^a,E^a_{\alpha})$ and $(\lambda_a,E_a^{\alpha})^{\mathrm{T}}$ are mutually 
inverse and since they are finite dimensional we also have
\begin{eqnarray}
	\begin{pmatrix}
	 \lambda_a\\
	 E_a^{\alpha}
	\end{pmatrix}
	\begin{matrix}
		\begin{pmatrix}
		\lambda^a & E^a_{\beta}
		\end{pmatrix}\\
		\begin{matrix}
		\end{matrix}
	\end{matrix}
=
\begin{pmatrix}
1 & 0 \\ 
0 & \delta^{\alpha}_{\beta}
\end{pmatrix}
,\end{eqnarray}
or
\begin{eqnarray}
	\begin{matrix}
		\,\,\lambda_a\lambda^a=1, & & \lambda_aE^a_{\alpha}=0,\\
		E_a^{\alpha}\lambda^a=0, & & \,E_a^{\alpha}E^{a}_{\beta}=\delta^{\alpha}_{\beta}.
	\end{matrix}
\end{eqnarray}
As we expected, variables $\lambda^a$, $\lambda_a$, $\EE^a$ and $\EE_a$ are not independent and we can 
express vector coefficients by using the covectors via well known formula for inverse matrix
\begin{eqnarray}
e\lambda_a&=&\frac{\partial \,e\,\,}{\partial\lambda^a},\\
e E^{\alpha}_a&=&\frac{\partial \,e\,\,}{\partial E^a_{\alpha}},
\end{eqnarray}
where 
\begin{eqnarray}
e=\frac{1}{3!}\bleps_{abcd}\bar{\varepsilon}^{\alpha\beta\gamma}\lambda^aE^b_{\alpha}E^c_{\beta}E^d_{\gamma}
\end{eqnarray}
is determinant of matrix $(\lambda^a,E^a_{\alpha})$. Coordinate's (co)vectors can be written with the help 
of previous formulas as
\begin{eqnarray}
	\begin{matrix}
		\dd t=\lambda_a\ee^a & & \dd x^{\alpha}=E^{\alpha}_a\ee^a\\
		\partial_t=\lambda^a\ee_a & & \,\,\partial_{\alpha}\,=E^a_{\alpha}\ee_a
	\end{matrix}\label{pr2}
\end{eqnarray}
thus we see that vector $\partial_t\in\Td^1\M$ is represented by vector $\lambda^a\in\Mkw$ and similar 
for $\dd t\in\Td_1\M$ we have $\lambda_a\in\Td_1\Mkw$. \\
\hs Since $\Mkw$ is isomorphic to $\Td^1\M$ and there exists a natural decomposition of $\Td^1\M$ into 
subspaces collinear with embedding of $\Sigma$ and $\partial_t$ there should also exist a similar structure 
on the space $\Mkw$. We have immediately from relation $(\lambda^a\lambda_c)(\lambda^c\lambda_b)=
\lambda^a\lambda_b$ that $\lambda^a\lambda_b$ is a projection on $\Mkw$. We can rearrange the equation 
(\ref{pr1}) as
\begin{eqnarray}
\EE^a_b=\EE^a(\EE_b)=E^a_{\alpha}E^{\alpha}_b=\delta^a_b-\lambda^a\lambda_b
\end{eqnarray}
and another supplemental projection $\EE^a_b$ on $\Mkw$ appears. It is clear from (\ref{pr2}) that 
$\lambda^a\lambda_b$ maps a general vector $v^a\in\Mkw$ on that part of $v^a$ which is proportional to
 $\partial_t$ and $\EE^a_b$ on that tangent to $\Sigma$.\\
\hs We were working with general orthonormal frame until now. From this moment $\ee^a$ is supposed to 
be righthanded and future oriented. This assumption restricts our variables $\lambda^a$, $\EE^a$ and 
following conditions should be fulfilled
\begin{eqnarray}
\lambda^0>0,&&\label{conf1}\\
\eta_{ab}\lambda^a\lambda^b>0,&&\\
e>0,&&\\
\q=\eta_{ab}\EE^a\otimes\EE^b<0,&&\label{conf4}
\end{eqnarray}
where $\q$ is spatial metric and $\q<0$ means that this tensor on $\Sigma$ is strictly negative, 
i.e $\forall \vv\neq 0\in\Td^1\Sigma:\q(\vv,\vv)<0$. Let $\SO(\g)$ be a subgroup of $\OOO(g)$ preserving
conditions (\ref{conf1})-(\ref{conf4}). If one wants to work with the whole $\OOO(\g)$ 
then configuration manifold splits into four disjoint 
parts given by future/past and right/left hand orientation and this discrete structure should be taken 
into account on quantum level, but this is far at the moment.\\
\hs Decomposition of variables $\hat{\BB}^{ab}$, $\hat{\CC}^{ab}$ is given by
\begin{eqnarray}
\hat{\BB}^{ab}&=&\B^{ab}\dd t+\BB^{ab},\\
\hat{\CC}^{ab}&=&\C^{ab}\dd t+\CC^{ab}
\end{eqnarray}
and we can now focus on the metric connection variable $\HAA^{ab}$. We can write 
\begin{eqnarray}
\HAA^{ab}=\Lambda^{ab}\dd t+\AAA^{ab}.
\end{eqnarray}
It should be noted that $\Lambda^{ab}$ transforms like tensor under $g\in\SO(\g)\times\Sigma$. Let $\tilde{\hat{\ee}}^a=g^*\hat{\ee}^a=O^a_{\phantom{a}b}\hat{\ee}^b$ be a new coframe\footnote{ $\,\,\ee^a$ is 
coframe on $\Td_1\M$, $\hat{\ee}^a$ is its representation on $\Td_1\Mkw$} on $\Td_1\Mkw$ then transformation 
law for $\AAA^{ab}$ is  given by formula
\begin{eqnarray}
\tilde{\AAA}^{ab}=O^a_{\phantom{a}{\bar{a}}}O^b_{\phantom{b}{\bar{b}}}\AAA^{\bar{a}\bar{b}}+
O^a_{\phantom{a}{\bar{a}}}\eta^{\bar{a}\bar{b}}\dd O^b_{\phantom{b}{\bar{b}}}.\nonumber 
\end{eqnarray}
Let $\hat{\vv}^a=\check{\vv}^a\wedge\dd t+\vv^a\in\Lambda\Td\M$ then $\HD\hat{\vv}^a$ can be written  as
\begin{eqnarray}
\HD\hat{\vv}^a=\D\check{\vv}^a\wedge\dd t+\dd t\wedge\dot{\D}\vv^a+\D\vv^a,
\end{eqnarray}
where $\D$ is a spatial covariant external derivative operator on $\Lambda\Td\Mkw$ given by
\begin{eqnarray}
\D\vv^a=\dd\vv^a+\eta_{bc}\AAA^{ab}\wedge\vv^c
\end{eqnarray}
and $\dot{\D}$ is a covariant time derivative
\begin{eqnarray}
\dot{\D}\vv^a=\dot{\vv}^a+\eta_{bc}\Lambda^{ab}\vv^c.
\end{eqnarray}
Since $\Lambda^{ab}$ and $\AAA^{ab}$ are antisymmetric in their indices we have immediately that
\begin{eqnarray}
\D\eta_{ab}=0
\end{eqnarray}
and
\begin{eqnarray}
\dot{\D}\eta_{ab}=0.
\end{eqnarray}
Thus operators $\D$ and $\dot{\D}$ are compatible with the metric $\eta_{ab}$ on $\Mkw$.\\
\hs Let us summarize our situation. We started with connection $\HD$ on $\Lambda\Td\M$ with gauge group $\SO(\g)$.
 $3+1$ decomposition
of space $\Lambda\Td\M$ leads us to the pure spatial connection $\D$ on $\Sigma$ with the same group $\SO(\g)$
 which is good news for us. Since as we wanted or expected the $\SO(\g)$ structure is preserved even in the 
language of spatial forms on $\Sigma$. This is in contrast with standard ADM/real Loop 
formulation\footnote{Of course ADM formalism works with spatial metric $\q$ and therefore there are no 
coframe variables. For example in the Loop gravity Hamiltonian formulation starts with ADM, then 
orthonormal coframe $\ee^i$ on $\Sigma$ is introduced and metric is expressed by orthonormality of 
this coframe, i.e. $\q$ ($i,j=1,2,3$).} where gauge group is only $\SO(\q)$. So far we are still working with real variables which 
is again in contrast with complex Loop theory where gauge group is $\SO(\g)$ but the prize paid for that is 
the loss of reality of variables.\\
\hs In general theory of gauge connections a notion of curvature is well known. Vanishing of curvature expresses
 the condition that a horizontal subspace in fibre bundle over given manifold is integrable. In usual 
words this means that parallel transport along closed path of a given object (the object should be valued in 
nontrivial representantion space of the gauge group) is given by identity (see details in ,e.g., [\cite{Fecko}]). 
That's why curvature plays important role even for general gauge group G (recall $\hat{\mathbf{F}}=
\hd \HAA$ in Maxwell theory or more complicated objects in Standard Model). For our purposes it is sufficient 
to write down an explicit formula which is 
\begin{eqnarray}
\RR^{ab}=\D\AAA^{ab}=\dd\AAA^{ab}+\eta_{cd}\AAA^{ac}\wedge\AAA^{db}\nonumber
\end{eqnarray}
for our $\SO(\g)$ connection $\AAA^{ab}$ on $\Lambda\Td\Mkw$. The spacetime curvature $\HR^{ab}$ can be 
decomposed as
\begin{eqnarray}
\HR^{ab}=\RR^{ab}+\dd t\wedge\dot{\AAA}^{ab}+\D\Lambda^{ab}\wedge\dd t.
\end{eqnarray}
\hs Next geometrical object on $\M$ which plays important role in Einstein-Cartan theory is the torsion 
$\hat{\TT}^a=\HD\ee^a$. How does its spatial counterpart look like? Coframe $\ee^a$ is not object from 
$\Lambda\Td\Mkw$ 
because it contains $\dd t$. We can project $\ee^a$ with $\EE^a_b$ and have $\EE^a=\EE^a_b\ee^b$ 
what is already an object from $\Lambda\Td\Mkw$. Thus, let us define $\SO(\g)$-torsion by formula
\begin{eqnarray}
\TT^a=\D\EE^a.
\end{eqnarray}
Since we are not and will not be working with the $3$-dimensional $\SO(\q)$-connection let us 
call for simplicity $\TT^a$ as torsion on places where no confusion can arise. Another motivation for its name 
appears if we write spacetime torsion $\hat{\TT}^a$ in $3+1$ manner
\begin{eqnarray}
\hat{\D}\ee^a=\D\EE^a+\D\lambda^a\wedge\dd t+\dd t\wedge\dot{\D}\EE^a.
\end{eqnarray}
As we can see, a spatial part of the spacetime torsion $\hat{\TT}^a$ is just $\SO(\g)$-torsion $\TT^a$.\\
\hs It will be useful in a while and also in next sections to have derived few formulas. In order to do this, 
let us consider 2-form $\PP_{ab}$ which is antisymmetric in its indices $ab$, i.e. 
\begin{eqnarray}
\PP_{ab}=\frac{1}{2}\tilde{P}_{ab}^{\alpha}\varepsilon_{\alpha\beta\gamma}\dd x^{\beta}\wedge\dd x^{\gamma}.
\end{eqnarray}
$\PP_{ab}$ can be decomposed in its tensor indices into spatial and time parallel parts as
\begin{eqnarray}
\PP_{ab}=2\PP^{^{\bot}}_{[a}\lambda^{\phantom{^{\bot}}}_{b]}+\hat{\PP}_{ab},
\end{eqnarray}
where 
\begin{eqnarray}
\PP^{^{\bot}}_a=\PP_{ab}\lambda^b,
\end{eqnarray}
note that $\PP^{^{\bot}}_a\lambda^a=0$, and
\begin{eqnarray}
\hat{\PP}_{ab}=\EE^{\bar{a}}_a\EE^{\bar{b}}_b\PP_{\bar{a}\bar{b}}.
\end{eqnarray}
Let us focus on the tangential part $\hat{\PP}_{ab}$. We can multiply it by $\EE^a$ 
\begin{eqnarray}
\KK^c_{ab}=\hat{\PP}_{ab}\wedge\EE^c\label{KK^c_ab}
\end{eqnarray}
It is easy to show that there is a one to one correspondence between $\hat{\PP}_{ab}$ and $\KK^c_{ab}$ iff $\lambda_c\KK^c_{ab}=0$, 
$\KK^c_{ab}=-\KK^c_{ba}$ and $\lambda^a\KK^c_{ab}=0$. Let $\KK^c_{ab}=
\tilde{K}^c_{ab}\dd^3 x$, then
$\tilde{K}^c_{ab}= \hat{\tilde{P}}^{\alpha}_{ab}E^c_{\alpha}$ and due to $\lambda_c\KK^c_{ab}=0$ we can express $\hat{\tilde{P}}^{\alpha}_{ab}=\tilde{K}^c_{ab}E_c^{\alpha}$. Equation (\ref{KK^c_ab}) can be rearranged
without any loss of information by multiplying with $\bar{\bleps}^{a\bar{b}\bar{c}\bar{d}}\lambda_{\bar{b}}$, since bottom 
indices are spatial and antisymmetric, into the 3-form
\begin{eqnarray}
\KK^{ab}=\frac{1}{2}\bar{\bleps}^{a\bar{b}\bar{c}\bar{d}}\lambda_{\bar{b}}\KK^b_{\bar{c}\bar{d}}=\frac{1}{2}
\bar{\bleps}^{a\bar{b}\bar{c}\bar{d}}\lambda_{\bar{b}}\PP_{\bar{c}\bar{d}}\wedge\EE^{b},
\end{eqnarray}
which can be written as a sum of symmetric and antisymmetric parts
\begin{eqnarray}
\KK^{ab}=\KK^{(ab)}+\KK^{[ab]}.
\end{eqnarray}
Antisymmetric part can be rewritten as
\begin{eqnarray}
\PP^{^{||}}_a=\bleps_{bcda}\lambda^b\KK^{[cd]}&=&-\frac{1}{2}\bleps_{abcd}\bar{\bleps}^{c\bar{b}\bar{c}\bar{d}}
\lambda^b\lambda_{\bar{b}}\hat{\PP}_{\bar{c}\bar{b}}\wedge\EE^d=\dots\nonumber\\
\PP^{^{||}}_a&=&\EE^b_a\PP_{bc}\wedge\EE^c.
\end{eqnarray}
Thus whole information about $\PP_{ab}$ is encoded in three independent components\\
\hs $\PP^{^{\bot}}_a$ $-$ 2-form spatial covector,\\
\hs $\PP^{^{||}}_a$ $\,-$ 3-form spatial covector,\\
\hs $\bsigma^{ab}$ $-$ spatial symmetric 3-form,\\
where (sign and 2 is just convention)
\begin{eqnarray}
\bsigma^{ab}=-2\KK^{(ab)}=\frac{1}{2}\PP_{\bar{a}\bar{b}}\lambda_{\bar{c}}\wedge
\left(\bar{\bleps}^{\bar{a}\bar{b}\bar{c} a}\EE^b+ \bar{\bleps}^{\bar{a}\bar{b}\bar{c} b}\EE^a\right).
\nonumber
\end{eqnarray}
Let us consider a linear map of $\PP_{ab}$ given by integral
\begin{eqnarray}
\PP(\BB)=\int\limits_{\Sigma}\frac{1}{2}\PP_{ab}\wedge\BB^{ab},
\end{eqnarray}
where $\BB^{ab}$ is a 1-form antisymmetric in its indices. Since we can decompose $\PP_{ab}$ into the three 
parts we can expect that similar decomposition works for its dual $\BB^{ab}$. We can write 
\begin{eqnarray}
\frac{1}{2}\PP_{ab}\wedge\BB^{^{\bot}ab}=\PP^{^{\bot}}_a\wedge\BB^a=\frac{1}{2}\PP_{ab}\wedge 
2\BB^{[a}\lambda^{b]},
\end{eqnarray}
thus $\BB^{^{\bot}ab}=2\BB^{[a}\lambda^{b]}$
\begin{eqnarray}
\frac{1}{2}\PP_{ab}\wedge\BB^{^{||}ab}=\PP^{^{||}}_a\B^{a}=\frac{1}{2}\PP_{ab}\wedge 2\B^{\bar{a}}\EE^{[a}_{\phantom{[}\bar{a}}\EE_{\phantom{\bar{a]}}}^{b]},
\end{eqnarray}
thus $\BB^{^{||}ab}=2\B^{\bar{a}}\EE^{[a}_{\phantom{[}\bar{a}}\EE^{b]}_{\phantom{a]}}$ and
\begin{eqnarray}
\frac{1}{2}\PP_{ab}\BB^{^Mab}=\bsigma^{ab}M_{ab}=\frac{1}{2}\PP_{ab}\lambda_{\bar{a}}
\bar{\bleps}^{ab\bar{a}\bar{b}}\wedge\EE^{\bar{c}}M_{\bar{b}\bar{c}},
\end{eqnarray}
thus $\BB^{^Mab}=\bar{\bleps}^{ab\bar{a}\bar{b}}\EE^{\bar{c}}\lambda_{\bar{a}}M_{\bar{b}\bar{c}}$. 
In other words we can decompose dual to $\PP_{ab}$ as
\begin{eqnarray}
\BB^{ab}=2\BB^{[a}\lambda^{b]}+2\B^{\bar{a}}\EE^{[a}_{\phantom{[}\bar{a}}\EE_{\phantom{\bar{b}]}}^{b]}+
\bar{\bleps}^{ab\bar{a}\bar{b}}\EE^{\bar{c}}\lambda_{\bar{a}}M_{\bar{b}\bar{c}},
\label{Rzkld_BB}
\end{eqnarray}
where $\BB^a$ is an arbitrary 1-form vector, $\B^a$ is a 0-form vector and $M_{ab}$ is a symmetric matrix.\\
\hs We already derived equations of motion of the Einstein-Cartan theory from Lagrangian in section 
\ref{Section-Lagrangian_of_Cartan_Theory}
 and now it is the right time to explore them in detail. Anyway, we present here only brief description 
and leave the rest to the next chapters where Hamiltonian-Dirac formalism is explored in full detail. 
Recall that the torsion equation (\ref{TorEQ}) sets
the connection to be just geometrical; in other words $\HAA^{ab}$, can be written as function(al) of the metric $g_{\mu\nu}=\eta_{ab}e^a_{\mu}e^{b}_{\nu}$ and 
initial value formulation for Einstein equations (\ref{EEQ}) 
written using $g_{\mu\nu}$
is well known and understood problem (see, e.g. \cite{Wald}). 
If we follow ideas of  Einstein-Cartan theory and work with our variables $\AAA^{ab}$, 
$\EE^a$, etc.
 then the set of equations given by (\ref{TorEQ}) and (\ref{EEQ}) is not complete. Missing equations should be 
derived from the condition preserving the constraints given by the equation (\ref{TorEQ}) and (\ref{EEQ}). 
Let us look what happens here. Decomposition of (\ref{TorEQ}) leads to
\begin{eqnarray}
&0&=\frac{1}{8\pi\kappa}\bleps_{abcd}\EE^c\wedge\D\EE^d,\label{EOM-3T}\\
&0&=\frac{1}{8\pi\kappa}\bleps_{abcd}\left(\lambda^c\D\EE^d+\EE^c\wedge\D\lambda^d-\EE^c\wedge\dot{\D}
\EE^d\right).
\label{EOM-3T-evolution}
\end{eqnarray}
Equations (\ref{EEQ}) can be rewritten similarly as
\begin{eqnarray}
&0&=-\frac{1}{16\pi\kappa}\bleps_{abcd}\RR^{bc}\wedge\EE^d\label{EOM-3R},\\
&0&=-\frac{1}{16\pi\kappa}\bleps_{abcd}\left(\RR^{bc}\lambda^d+\dot{\AAA}^{bc}\wedge\EE^d-\D\Lambda^{bc}.
\wedge\EE^d\right)\label{EOM-3R-evolution}
\end{eqnarray}
The expression on the right-hand side of (\ref{EOM-3T-evolution}) is a 2-form with antisymmetric indices 
and we can use decompositon (\ref{Rzkld_BB}).
We obtain an evolution equation and a constraint
\begin{eqnarray}
&0&=\dot{\D}\EE^a-\D\lambda^a,\label{EOM-DdotE}\\
&0&=\EE^{(a}\EE^{b)}_{c}\wedge\D\EE^c.\label{EOM-sym3T}
\end{eqnarray}
Here is no problem with ambiguity. The equation (\ref{EOM-3R-evolution}) is a 2-form with one 
tensor index hence it expresses $4\times 3=12$ conditions for $\dot{\AAA}^{ab}$ with $6\times 3=18$ degrees of 
freedom. We
see that we are not able to determine connection velocities and some equation(s) is(are) still missing.
We will see later that conditions (\ref{EOM-3T}) and (\ref{EOM-3R}) represent the first class contraints while
equation (\ref{EOM-sym3T}) is the constraint of the second class. Missing equation(s) can be obtained by 
applying the time derivative on (\ref{EOM-sym3T}). Since (\ref{EOM-3R}) and (\ref{EOM-3T}) are the first 
class constraints
no new conditions appear and we have closed system of equations determining $\EE^a$ and $\AAA^{ab}$. 
The variables $\lambda^a$ and $\Lambda^{ab}$ are arbitrary. The missing equation is 
\begin{eqnarray}
0=\EE^{(a}\EE^{b)}_c\wedge(\RR^{c\bar{a}}\eta_{\bar{a}\bar{b}}\lambda^{\bar{b}}+\HH^{c\bar{a}}
\eta_{\bar{a}\bar{b}}\wedge\EE^{\bar{b}}),
\end{eqnarray}
where $\HH^{ab}=\dot{\AAA}^{ab}-\D\Lambda^{ab}$. Now we can determine $\HH^{ab}$ as a certain function(al)
of $\lambda^a$, $\EE^a$, $\AAA^{ab}$ but we will not do that because we do not need it anywhere. It is enough 
for our purposes to know that our set of equations determines uniquely, up to gauge transformation, evolution 
of our system.
%%%%%%%%%%%%%%%%%%%%%%%%%%%%%%%%%%%%%%%%%%%%%%%%%%%%%%%%%%%%%%%%%%%%%%%%%%%%%%%%%%%%%%%%%%%%%%
%%%%%%%%%%%%%%%%%%%%%%%%%%%%%  Hamiltonian %%%%%%%%%%%%%%%%%%%%%%%%%%%%%%%%%%%%%%%%%%%%%%%%%%%
%%%%%%%%%%%%%%%%%%%%%%%%%%%%%%%%%%%%%%%%%%%%%%%%%%%%%%%%%%%%%%%%%%%%%%%%%%%%%%%%%%%%%%%%%%%%%%

\section{Hamiltonian}\label{section-Hamiltonian}

In section \ref{Section-Lagrangian_of_Cartan_Theory} we have introduced the Lagrangian of  
the Einstein-Cartan theory. Next step towards its quantum formulation should be done by its 
converting it into canonical form. Since our system contains velocities of basic 
variables at best linearly, standard Hamilton procedure can not be used. Therefore we must 
use Dirac procedure for constrained dynamic \cite{Dirac}. In the standard and even in the Dirac approach to 
dynamics the notion of momentum for variable $q^A$ is introduced by $p_A=\frac{\partial L}
{\partial\dot{q}^A}$, 
where $L$ is Lagrangian of a system. Since action is $S=\int\dd t L$ we can see that action and 
Lagrangian
for field theory can be written within 4-form $\LL$ called
Lagrangian form as $S=\int\limits_{\Omega}\LL$ and $L=\int\limits_{\Sigma}i_{\partial_t}\LL$,
where $\LL=\Lg\dd^4x$ and $\Lg$ is Lagrangian density. If we suppose 
that configuration space is built just by generalized $n$-forms 
$\mathbf{Q}^A=\frac{1}{n!}Q^A_{\alpha\dots\beta}\dd x^{\alpha}\wedge\dots\wedge\dd x^{\beta}$, e.g. $\EE^a$, $\AAA^{ab}$ in our system, 
all variables in Standard Model, etc., then we can see that their momenta
$\tilde{p}^{\alpha\dots\beta}_A=\frac{\delta L}{\delta\dot{Q}^A_{\alpha\dots\beta}}=\frac{\partial\Lg}{\partial\dot{Q}^A_{\alpha\dots\beta}}$ transform 
like densities under coordinate transformation and therefore objects
$\pp_A=\frac{1}{n!(3-n)!}\tilde{p}^{\alpha\dots\beta}_A\varepsilon_{\alpha\dots\beta\gamma\dots\delta}\dd x^{\gamma}\wedge\dots\wedge\dd x^{\delta}$ are $(3-n)$-forms and even more 
$\pp_A\wedge\dot{\mathbf{Q}}^A=\frac{1}{n!}\tilde{p}^{\alpha\dots\beta}_A\dot{Q}^A_{\alpha\dots\beta}\dd^3 x$ 
what is exactly the first term in the definition of Hamiltonian 
$H=\int\limits_{\Sigma}\pp_A\wedge\dot{\mathbf{Q}}^A-L$. Recall that $Q^A_{\alpha\dots\beta}$ and
$\tilde{p}^{\alpha\dots\beta}_A$ are antisymmetric in their coordinate indices therefore every term
in $\tilde{p}^{\alpha\dots\beta}_A\dot{Q}^A_{\alpha\dots\beta}$ is $n!$-times repeated 
while every velocity should enter the Hamiltonian just once. Our configuration space is described by variables 
$\lambda^a, \dots, \CC^{ab}$ and its velocities (see table \ref{tabulka} for details). Variables 
$\B,\dots,\CC^{ab}$ enters the Lagrangian (\ref{3+1-Lagrangian}) in a certain special way.
We can decompose it as sum of two Lagrangians $\LL=\LLC+\LLR$ where 
\begin{table}[t]
\caption{Basic variables}\label{tabulka}
\begin{center}
    \begin{tabular}{| l | l | l |}
    \hline
    Variables & Momenta & Velocities \\ \hline
    $\lambda^a$ 
	& 
	 $\blpi_a=\tilde{\pi}_a\dd^3 x$ where $\tilde{\pi}_a=\partial \Lg / \partial\dot{\lambda}^a$ 
	& 
	 $\nu^a=\dot{\lambda}^a$
   \\ 
    $\EE^a=E^a_{\alpha}\dd x^{\alpha} $  
	&
	 $\pp_a=\frac{1}{2}\tilde{p}_a^{\alpha}\varepsilon_{\alpha\beta\gamma}\dd x^{\beta}\wedge\dd x^{\gamma}$ where 
		$\tilde{p}_a^{\alpha}=\partial \Lg / \partial\dot{E}^a_{\alpha}$ 
	& 
		$\bb^a=\dot{\EE}^a$
   \\ 
    $\Lambda^{ab}$ 
	& 
	 $\blPI_{ab}=\tilde{\Pi}_{ab}\dd^3 x$ where $\tilde{\Pi}_{ab}=\partial \Lg / \partial\dot{\Lambda}^{ab}$
	& 
	 $\Gamma^{ab}=\dot{\Lambda}^{ab}$
	\\ 
    $\AAA^{ab}=A^{ab}_{\alpha}\dd x^{\alpha}$ 
	& 
	 $\pp_{ab}=\frac{1}{2}\tilde{p}_{ab}^{\alpha}\varepsilon_{\alpha\beta\gamma}\dd x^{\beta}\wedge\dd x^{\gamma}$ where
	 $\tilde{p}_{ab}^{\alpha}=\partial \Lg / \partial\dot{A}^{ab}_{\alpha}$ 
	& 
	 $\BB^{ab}=\dot{\AAA}^{ab}$ 
	\\
	 $\B$ 
	&
	 $\blphi=\tilde{\varphi}\dd^3 x$ where $\tilde{\varphi}=\partial\Lg / \partial\dot{\B}$
	&
	 $\Y=\dot{\B}$
	\\ 
	 $\BB=B_{\alpha}\dd x^{\alpha}$
	&
	 $\uu=\frac{1}{2}\tilde{u}^{\alpha}\varepsilon_{\alpha\beta\gamma}\dd x^{\beta}\wedge\dd x^{\gamma}$ where 
		$\tilde{u}^{\alpha}=\partial\Lg / \partial\dot{B_{\alpha}}$
	&
	 $\YY=\dot{\BB}$
	\\ 
	 $\C^{ab}$
	&
	 $\blPHI_{ab}=\tilde{\Phi}_{ab}\dd^3 x$ where 
		$\tilde{\Phi}_{ab}=\partial\Lg / \partial\dot{\C^{ab}}$
	&
	 $\X^{ab}=\dot{\C^{ab}}$
	\\ 
	 $\CC^{ab}=C^{ab}_{\alpha}\dd x^{\alpha}$
	&
	 $\UU_{ab}=\frac{1}{2}\tilde{U}_{ab}^{\alpha}\varepsilon_{\alpha\beta\gamma}\dd x^{\beta}\wedge\dd 
x^{\gamma}$ where 
		$\tilde{U}^{\alpha}_{ab}=\partial\Lg / \partial\dot{C}^{ab}_{\alpha}$
	&
	 $\XX^{ab}=\dot{\CC}^{ab}$
\\ \hline
    \end{tabular}
\end{center}

\end{table}
\begin{eqnarray}
\LLR=-\dd t\wedge\frac{1}{16\pi\kappa}\eta_{\bar{a}\bar{b}}\bleps_{abcd}
		(\C^{a\bar{a}}\CC^{b\bar{b}}\wedge\EE^c\wedge\EE^d+\CC^{a\bar{a}}\wedge\CC^{b\bar{b}}\wedge\lambda^c\EE^d)
\end{eqnarray}
and $\LLC$ does not depend on $\C^{ab}$, $\CC^{ab}$ while as we already know, the whole Lagrangian $\LL$ does 
not depend on $\B$, $\BB$. Thus we can consider this subsystem independently. Hamiltonian $\HHR$ is given by
\begin{eqnarray}
&&\HHR=\phantom{+\,}\blphi\wedge\Y+\uu\wedge\YY+\frac{1}{2}\blPHI_{ab}\wedge\X^{ab}+\frac{1}{2}\UU_{ab}
\wedge\XX^{ab}+\nonumber\\
&&\phantom{\HHR=} +\frac{1}{16\pi\kappa}\eta_{\bar{a}\bar{b}}\bleps_{abcd}
		(\C^{a\bar{a}}\CC^{b\bar{b}}\wedge\EE^c\wedge\EE^d+\CC^{a\bar{a}}\wedge\CC^{b\bar{b}}\wedge\lambda^c\EE^d)
\end{eqnarray}
with primary constraints $\blphi=\uu=\blPHI_{ab}=\UU_{ab}=0$. Secondary constraints are $\C^{ab}=\CC^{ab}=0$.
Since the constraints $\blPHI_{ab}$, $\UU_{ab}$ and $\C^{ab}$, $\CC^{ab}$ are mutually canonically conjugated,  their Poisson bracket is an identity, they are the second class constraints and we must use the Dirac procedure. Dirac bracket for this subsystem is just Poisson bracket on canonical variables $\B$, $\BB$ and momenta $\blphi$, $\uu$ while reduced Hamiltonian is $ \HHR=\blphi\wedge\Y+\uu\wedge\YY $. Hence we can focus ourselves for a while just on $\LLC$ and its hamiltonization.  Final Hamiltonian will be obtained by sum 
$\HH=\HHC+\HHR$.\\
\hs Let us substitute the decomposition of the variables $\ee^a$, $\HAA^{ab}$ into the Langrangian $\LLC$
\begin{eqnarray}
i_{\partial_t}\LLC&=&
-\frac{1}{16\pi\kappa} \bleps_{abcd}\lambda^{a}\RR^{bc}\wedge\EE^d+ \frac{1}{32\pi\kappa}\bleps_{abcd}\D\Lambda^{ab}\wedge\EE^c\wedge\EE^d\nonumber\\ &\phantom{=}&-\frac{1}{32\pi\kappa}\bleps_{abcd}\dot{\AAA}^{ab}\wedge\EE^c\wedge\EE^d.\label{3+1-Lagrangian}
\end{eqnarray}
We use this in definition of Hamiltonian. Our procedure then yields the following result
\begin{eqnarray}
\ham=\int\limits_{\Sigma}\HHC=\blpi(\nu)+\blPI(\Gamma)+\pp(\bb)+\PP(\BB)+\RR(\lambda)+\TT(\Lambda),
\end{eqnarray}
where
\begin{eqnarray}
&\blpi(\nu)&=\int\limits_{\Sigma}\blpi_a\wedge\nu^a,\nonumber\\
&\pp(\bb)&=\int\limits_{\Sigma}\pp_a\wedge\bb^a,\nonumber\\
&\blPI(\Gamma)&=\int\limits_{\Sigma}\frac{1}{2}\blPI_{ab}\wedge\Gamma^{ab},\nonumber\\
&\PP(\BB)&=\int\limits_{\Sigma}\frac{1}{2}\left(\pp_{ab}+\frac{1}{16\pi\kappa}
					\bleps_{abcd}\EE^c\wedge\EE^d\right)\wedge\BB^{ab}=
	\int\limits_{\Sigma}\frac{1}{2}\PP_{ab}\wedge\BB^{ab},\nonumber\\
&\RR(\lambda)&=\int\limits_{\Sigma}\frac{1}{16\pi\kappa}\bleps_{abcd}\lambda^a\RR^{bc}
					\wedge\EE^d=\int\limits_{\Sigma}\lambda^a\RR_a,\nonumber\\
&\TT(\Lambda)&=\int\limits_{\Sigma}\!\!-\frac{1}{32\pi\kappa}\bleps_{abcd}\D\Lambda^{ab}
					\wedge\EE^c\wedge\EE^d=
\int\limits_{\Sigma}-\frac{1}{16\pi\kappa}\bleps_{abcd}\Lambda^{ab}\wedge\EE^c\wedge\D\EE^d=
\int\limits_{\Sigma}\frac{1}{2}\Lambda^{ab}\TT_{ab},\nonumber
\end{eqnarray}
where $\PP_{ab}=\pp_{ab}+\frac{1}{16\pi\kappa}\bleps_{abcd}\EE^c\wedge\EE^d$, 
$\RR_a=\frac{1}{16\pi\kappa}\bleps_{abcd}\RR^{bc}\wedge\EE^d$ and 
\\$\TT_{ab}=-\frac{1}{8\pi\kappa}\bleps_{abcd}\EE^c\wedge\D\EE^d$.
The existence of the primary constraints represents the fact that we are working with a degenerated Lagrangian 
and therefore we are not able to express velocities as function(al)s of momenta (they are given by conditions 
$\frac{\partial\Lg}{\partial{\dot{Q}^A}}=0$). Our system is degenerated and primary contraints are given
by  
\begin{eqnarray}
\blpi(\nu)= 0&\,\,\,\,\,\, \forall\nu^a\in\Lambda_0\Td\Mkw
				&\Leftrightarrow\,\,\,\blpi_a=0,\nonumber\\
\pp(\bb)= 0&\,\,\,\,\,\, \forall\bb^a\in\Lambda_1\Td\Mkw
				&\Leftrightarrow\,\,\,\pp_a=0,\nonumber\\
\blPI(\Gamma)= 0&\,\,\,\,\,\, \forall\Gamma^{ab}\in\Lambda_0\Td\Mkw
				&\Leftrightarrow\,\,\,\blPI_{ab}=0,\nonumber\\
\PP(\BB)= 0& \,\,\,\,\,\, \forall\BB^{ab}\in\Lambda_1\Td\Mkw
				&\Leftrightarrow\,\,\,\PP_{ab}=\pp_{ab}+\frac{1}{16\pi\kappa}\bleps_{abcd}\EE^c\wedge\EE^d=0.
\nonumber
\end{eqnarray}
Since these constraints should be valid through the whole time evolution of our physical system their 
time derivatives should vanish too and this implies further conditions which should be fulfilled\footnote{
We omitted writing of details like $\forall\tilde{\nu}^a\dots$ in constraint's expressions.},
\begin{eqnarray}
\!\!\!\!\!\!\!\!\!\!\!\!\!\!\!\!\!\!\!\!\!\!
\frac{\dd \blpi(\tilde{\nu})}{\dd t}&=&
	\left\{\blpi(\tilde{\nu});\ham \right\}=-\RR(\tilde{\nu})=0,\label{RR-vazba}\\
\!\!\!\!\!\!\!\!\!\!\!\!\!\!\!\!\!\!\!\!\!\!
\frac{\dd \blPI(\tilde{\Gamma})}{\dd t}&=&
	\left\{\blPI(\tilde{\Gamma});\ham \right\}=-\TT(\tilde{\Gamma})=0,\label{TT-vazba}\\
\!\!\!\!\!\!\!\!\!\!\!\!\!\!\!\!\!\!\!\!\!\!
\frac{\dd \pp(\tilde{\bb})}{\dd t}&=&
	\left\{\pp(\tilde{\bb});\ham\right\}=\nonumber\\
	&=&
\int\frac{1}{16\pi\kappa}\bleps_{abcd}\tilde{\bb}^a\wedge\left(
\BB^{bc}\wedge\EE^{d}+\lambda^b\RR^{cd}-\D\Lambda^{bc}\wedge\EE^d
\right)=0,\label{POB-3R-evolution}\\
\!\!\!\!\!\!\!\!\!\!\!\!\!\!\!\!\!\!\!\!\!\!
\frac{\dd \PP(\tilde{\BB})}{\dd t}&=&
\left\{\PP(\tilde{\BB});\ham\right\}=\nonumber\\
&=&\int\frac{1}{16\pi\kappa}\bleps_{abcd}\tilde{\BB}^{ab}\wedge\left(
\bb^c\wedge\EE^d+\eta_{\bar{a}\bar{b}}\Lambda^{c\bar{a}}\EE^{\bar{b}}\wedge\EE^{d}-\D(\lambda^{c}\EE^{d})
\right)=0.\label{POB-3T-evolution}
\end{eqnarray}
The first two of them are secondary constraints. It is clear that (\ref{POB-3R-evolution}) is equal to
(\ref{EOM-3R-evolution}), while (\ref{POB-3T-evolution}) is connected with (\ref{EOM-3T-evolution});
they determine Lagrange multipliers $\bb^a$, $\BB^{ab}$. As we have already
promised in the previous section we will show how to do this now. Since these equations are same one can also use 
the same procedure there (recall that $\bb^a=\dot{\EE}^a$ and $\BB^{ab}=\dot{\AAA}^{ab}$). We can express 
the equations (\ref{POB-3R-evolution}), (\ref{POB-3T-evolution}) as:
\newcommand{\hh}{\mathbf{h}}
\begin{eqnarray}
0&=&\frac{1}{16\pi\kappa}\bleps_{abcd}(\HH^{bc}\wedge\EE^{d}+\RR^{bc}\lambda^{d}),\label{prva}\\
0&=&\frac{1}{8\pi\kappa}\bleps_{abcd}(\hh^{c}\wedge\EE^{d}-\lambda^{c}\D\EE^d),\label{druha}
\end{eqnarray}
where $\HH^{ab}=\BB^{ab}-\D\Lambda^{ab}$ and 
$\hh^a=\bb^a+\Lambda^{a\bar{a}}\eta_{\bar{a}\bar{b}}\EE^{\bar{b}}-\D\lambda^a$. Let us focus on the second
equation (\ref{druha}). We can multiply it again by a general 1-form $\tilde{\BB}^{ab}$ and since it is 
antisymmetric in its indices we can decompose it as (\ref{Rzkld_BB})
\begin{eqnarray}
\frac{1}{8\pi\kappa}(\tilde{\BB}^a\lambda^b+\tilde{\B}^{\bar{a}}\EE^a_{\bar{a}}\EE^b+
\frac{1}{2}\bar{\bleps}^{ab\bar{a}\bar{b}}\EE^{\bar{c}}\lambda_{\bar{a}}\tilde{M}_{\bar{b}\bar{c}})\wedge
\bleps_{abcd}(\hh^c\wedge\EE^d-\lambda^c\D\EE^d)=0.
\end{eqnarray}
This expression can be split into three independend equations
\begin{eqnarray}
\frac{1}{8\pi\kappa}\bleps_{abcd}\lambda^b\hh^c\wedge\EE^d=0,\\
\frac{1}{8\pi\kappa}\bleps_{abcd}\EE^{a}_{\bar{a}}\EE^b\wedge(\hh^c\wedge\EE^d-\lambda^c\D\EE^d)=0,\\
-\frac{1}{8\pi\kappa}\EE^{(a}_{\phantom{c}}\EE^{b)}_c\wedge\D\EE^c=0.
\end{eqnarray}
We can use constraint $\TT_{ab}=0$ in the second equation which together with the first one implies
that $\hh^a=0$, while the third equation is equivalent to another secondary constraints,
\newcommand{\SSS}{\mathbf{S}}
\begin{eqnarray}
\SSS(M)=\int\limits_{\Sigma}\frac{1}{8\pi\kappa}M_{ab}\EE^{a}_{\phantom{c}}\EE^{b}_c\wedge\D\EE^c
=\int\limits_{\Sigma}M_{ab}\SSS^{ab}=0,\label{SS-vazba}
\end{eqnarray}
where $\SSS^{ab}=\frac{1}{8\pi\kappa}\EE^{(a}_{\phantom{c}}\EE^{b)}_c\wedge\D\EE^c$ and $M_{ab}$ is arbitrary
function symmetric in its indices. Let us substitute the decomposition
\begin{eqnarray}
\HH^{ab}=2\HH^{[a}\lambda^{b]}+2\mathcal{H}^{\bar{a}}\EE_{\,\bar{a}}^{[a}\EE_{\phantom{\bar{a}}}^{b]}+
\bar{\bleps}^{ab\bar{a}\bar{b}}\EE^{\bar{c}}\lambda_{\bar{a}}\gamma_{\bar{b}\bar{c}}
\end{eqnarray}
into the equation (\ref{prva}) (where $\gamma_{ab}=\gamma_{ba}$). We obtain
\begin{eqnarray}
\frac{1}{16\pi\kappa}\bleps_{abcd}(2\HH^b\lambda^c\wedge\EE^d
	+2\mathcal{H}^{\bar{b}}\EE_{\bar{b}}^{b}\EE^{c}\wedge\EE^d
	+\RR^{bc}\lambda^d)=0
\end{eqnarray}
and if we multiply it with $\lambda^a$ then we have immediately that $\EE^a_b\mathcal{H}^{b}=0$ while 
$\lambda_a\mathcal{H}^{a}$ is arbitrary but we do not need it since it does not enter $\HH^{ab}$.
Hence this equation is reduced as
\begin{eqnarray}
\frac{1}{16\pi\kappa}\bleps_{abcd}(2\HH^b\lambda^c\wedge\EE^d
	+\RR^{bc}\lambda^d)=0,
\end{eqnarray}
which can be rewritten after some algebraic manipulations as
\begin{eqnarray}
2H^{[a}_{\,d}\lambda^{b]}_{\phantom{\,d}}+
2H^c_c\delta^{[a}_{\,d}\lambda^{b]}_{\phantom{\,d}}=-2R^{c[a}_{\phantom{c\,a}cd}\lambda^{b]},
\end{eqnarray}
where $R^{ab}_{\phantom{ab}cd}=i_{\EE_d}i_{\EE_c}\RR^{ab}$ and $H^a_b=i_{\EE_b}\HH^a$. 
Constraint $\RR_a\lambda^a=0$ is equivalent to $R^{ab}_{\phantom{ab}ab}=0$ and if we sum in previous
equation over $a=d$ then $H^a_a=0$ and we finally have 
\begin{eqnarray}
2\HH^{[a}\lambda^{b]}=-2i_{\EE_c}\RR^{c[a}\lambda^{b]}
\end{eqnarray}
or
\begin{eqnarray}
\HH^{ab}=-2i_{\EE_c}\RR^{c[a}\lambda^{b]}
			+\bar{\bleps}^{ab\bar{a}\bar{b}}\EE^{\bar{c}}\lambda_{\bar{a}}\gamma_{\bar{b}\bar{c}},
\label{HH-vyjadrenie_cez_gamma}
\end{eqnarray}
where $\gamma_{ab}$ is not determined yet. But there is no need to worry since our analysis 
is not over. We have just finished the first level of the Dirac procedure, however conservation of the 
secondary constraints should be analyzed too and there will appear the missing equation for $\gamma_{ab}$. 
In order to do this let us compute time derivatives of secondary constraints 
(\ref{RR-vazba}), (\ref{TT-vazba})
and (\ref{SS-vazba})
\begin{eqnarray}
\!\!\!\!\!\!
\frac{\dd \RR(\mu)}{\dd t}&=&\left\{\RR(\mu),\ham\right\}=\nonumber\\
	&=&\int\limits_{\Sigma}
				\frac{1}{16\pi\kappa}\bleps_{abcd}\,\mu^a(\RR^{bc}\wedge\bb^d+\D\BB^{bc}\wedge\EE^d)
	=0,\label{RR-der}\\
\!\!\!\!\!\!
\frac{\dd \TT(\Theta)}{\dd t}&=&\left\{\TT(\Theta),\ham\right\}=\nonumber\\
	&=&\int\limits_{\Sigma}\!\!\!-\frac{1}{16\pi\kappa}\bleps_{abcd}\Theta^{ab}
	(\EE^c\wedge\D\bb^d+\EE^c\wedge\BB^{d\bar{a}}\eta_{\bar{a}\bar{b}}\wedge\EE^{\bar{b}})
	=0,\label{TT-der}\\
\!\!\!\!\!\!
\frac{\dd \SSS(M)}{\dd t}&=&\left\{\SSS(M),\ham\right\}=\nonumber\\
	&=&\int\limits_{\Sigma}\frac{1}{8\pi\kappa}M_{ab}
\left(\EE^a\EE^b_c\wedge\D\bb^c
		+\EE^a\EE^b_c\wedge\BB^{c\bar{a}}\eta_{\bar{a}\bar{b}}\wedge\EE^{\bar{b}}\right)=0,
\label{SSS-der}
\end{eqnarray}
where the terms obviously proportional to the constraints are omitted. We can substitute from 
$\hh^a=\bb^a+\Lambda^{a\bar{a}}\eta_{\bar{a}\bar{b}}\EE^{\bar{b}}=0$ the expression for $\bb^a$ into 
(\ref{RR-der}) and thanks to generalized Bianchi $\D\RR^{ab}=0$ and Ricci 
$\D\D\Lambda^{ab}=\RR^{a\bar{a}}\eta_{\bar{a}\bar{b}}\Lambda^{\bar{a}b}+\RR^{b\bar{a}}\eta_{\bar{a}\bar{b}}\Lambda^{a\bar{b}}$
identities we have immediately
\begin{eqnarray}
\D\left(\frac{\bleps_{abcd}}{16\pi\kappa}(\RR^{bc}\lambda^{d}+\HH^{bc}\wedge\EE^d)\right)
\!-\!\frac{\bleps_{abcd}}{16\pi\kappa}\RR^{bc}\Lambda^{d\bar{a}}\eta_{\bar{a}\bar{b}}\EE^{\bar{b}}
\!\!+\!\!\frac{\bleps_{abcd}}{8\pi\kappa}\RR^{b\bar{a}}\eta_{\bar{a}\bar{b}}\Lambda^{\bar{b}c}\wedge\EE^d=0.
\nonumber
\end{eqnarray}
The first term vanishes due to (\ref{prva}). The last term can be transformed with the help of identity 
$\RR^{ab}=\frac{1}{4}\bar{\bleps}^{ab\bar{a}\bar{b}}\bleps_{\bar{a}\bar{b}\bar{c}\bar{d}}\RR^{\bar{c}\bar{d}}$
into expression
\begin{eqnarray}
\frac{\bleps_{abcd}}{8\pi\kappa}\RR^{b\bar{a}}\eta_{\bar{a}\bar{b}}\Lambda^{\bar{b}c}\wedge\EE^d=
\frac{\bleps_{abcd}}{16\pi\kappa}\RR^{bc}\Lambda^{d\bar{a}}\eta_{\bar{a}\bar{b}}\EE^{\bar{b}}-
\frac{1}{16\pi\kappa}\eta_{ab}\Lambda^{b\bar{a}}\bleps_{\bar{a}\bar{b}\bar{c}\bar{d}}
\RR^{\bar{b}\bar{c}}\EE^{\bar{d}}.\nonumber
\end{eqnarray}
Hence  no new condition appears from equation (\ref{RR-der}) since last term is proportional to $\RR_a=0$.\\
\hs Equation (\ref{TT-der}) can be rewritten with help $\hh^a=0$ and due to the fact that constraints 
$\TT_{ab}=\SSS^{ab}=0$ imply $\D\EE^a=0$ as
\begin{eqnarray}
\frac{1}{32\pi\kappa}\bleps_{abcd}\RR^{cd}\wedge\EE^{\bar{a}}\eta_{\bar{a}\bar{b}}\lambda^{\bar{b}}
+\frac{1}{8\pi\kappa}\eta_{\bar{a}[a}\bleps_{b]c\bar{b}\bar{c}}\EE^{\bar{a}}\wedge 
i_{\EE_{\bar{d}}}\RR^{\bar{d}\,\bar{b}}\lambda^{\bar{c}}\wedge\EE^c=0,\label{TT-der-po_uprave}
\end{eqnarray}
where (\ref{HH-vyjadrenie_cez_gamma}) has been already substituted. Since any 4-form on the three-dimensional
manifold vanishes identically we have that $\EE^a\wedge\RR^{bc}\wedge\EE^d=0$. We can apply interior product
on it $i_{\EE_b}(\EE^a\wedge\RR^{bc}\wedge\EE^d)=\EE^a_b\RR^{bc}\wedge\EE^d
-\EE^a\wedge i_{\EE_b}\RR^{bc}\wedge\EE^d-\EE^a\wedge\RR^{bc}\EE^d_b=0$. Now we can express 
from this the term proportional to $i_{\EE_b}\RR^{bc}$ and substitute it into previous equation. If
we use again 
$\RR^{ab}=\frac{1}{4}\bar{\bleps}^{ab\bar{a}\bar{b}}\bleps_{\bar{a}\bar{b}\bar{c}\bar{d}}\RR^{\bar{c}\bar{d}}$
then we finally find out that (\ref{TT-der-po_uprave}) is proportional to $\D\EE^a$. Hence again no new 
constraint
appears from (\ref{TT-der}).\\
\hs Equation (\ref{SSS-der}) can be rewritten as
\begin{eqnarray}
\frac{1}{8\pi\kappa}\EE^{(a}_{\phantom{c}}\EE^{b)}_c\wedge(\RR^{c\bar{a}}\eta_{\bar{a}\bar{b}}\lambda^{\bar{b}}+
\HH^{c\bar{a}}\eta_{\bar{a}\bar{b}}\wedge\EE^{\bar{b}})=0.\label{SSS-der-vyjadrenie}
\end{eqnarray}
This is the equation which determines $\gamma_{ab}$ entering (\ref{HH-vyjadrenie_cez_gamma}). 
However, we do not need explicit expression. For our purposes it is sufficient to show that this equation determines 
$\gamma_{ab}$ uniquely. In order to see it we should substitute the expression (\ref{HH-vyjadrenie_cez_gamma}) 
instead of $\HH^{ab}$ into this equation. Since (\ref{SSS-der-vyjadrenie}) is linear in $\HH^{ab}$ it is 
also linear in $\gamma_{ab}$, i.e. $c_A+Q_A^B\gamma_B$=0, where $A,B=(ab)$, and hence it is sufficient 
to show that $Q^A_B$ is invertible. The first observation is that (\ref{SSS-der-vyjadrenie}) actually 
represents
$6$ equations for $6$ pieces $\EE^a_{\bar{a}}\EE^b_{\bar{b}}\gamma_{ab}$ hence we can consider only the term 
proportional to $\gamma_{ab}$  which is $\lambda_b\lambda_{\bar{b}}\eta_{c\bar{c}}
\bar{\bleps}^{dcb(a}\bar{\bleps}^{\bar{a})\bar{b}\bar{c}\bar{d}}\gamma_{d\bar{d}}
=\tilde{G}^{a\bar{a}b\bar{b}}\gamma_{b\bar{b}}$ and as we will see in the next
section the expression $\tilde{G}^{a\bar{a}b\bar{b}}$ standing before $\gamma_{b\bar{b}}$ is invertible on
spatial subspace.\\
\hs Let us summarize this section. We have built the Hamiltonian formulation of  Einstein-Cartan theory.
The Hamiltonian is given by the sum of two Hamiltonians 
\newcommand{\hamcel}{\mathsf{H}}
\newcommand{\hamres}{\mathsf{H}^{\text{(Rest)}}}
\begin{eqnarray}
\!\!\!\!\!\!\!\!\!
\hamcel=\ham\!+\!\hamres\!&=&\blpi(\nu)\!+\!\blPI(\Gamma)\!+\!\pp(\bb)\!+\!\PP(\BB)\!
+\!\RR(\mu)\!+\!\TT(\Theta)\!+\!\SSS(M)\nonumber\\
						  &\phantom{=}&+\blphi(\Y)\!+\!\uu(\YY).
\end{eqnarray}
Constraints given by $\blpi(\nu)$, $\blPI(\Gamma)$, $\RR(\mu)$, $\TT(\Theta)$, $\blphi(\Y)$ 
and $\uu(\YY)$ do not determine any Lagrange multipliers, therefore they are the first class constraints.
The remaining constraints $\pp(\bb)$, $\PP(\BB)$ and $\SSS(M)$ are of the second class. Lagrange multipliers
$\bb^a$ and $\BB^{ab}$ are
\begin{eqnarray}
\bb^a\,\,&=&\D\lambda^a-\Lambda^{a\bar{a}}\eta_{\bar{a}\bar{b}}\EE^{\bar{b}},\\
\BB^{ab}&=&\D\Lambda^{ab}+\HH^{ab},
\end{eqnarray}
where $\HH^{ab}$ does not depend on $\Lambda^{ab}$ and it is the solution of (\ref{prva}) 
and (\ref{SSS-der-vyjadrenie}). We will continue with Dirac analysis in the next section where we will 
introduce Dirac bracket and explore the reduced phase space of our physical system.

%%%%%%%%%%%%%%%%%%%%%%%%%%%%%%%%%%%%%%%%%%%%%%%%%%%%%%%%%%%%%%%%%%%%%%%%%%%%%%%%%%%%%%%%%%%%%%%%%%%%%%%%%%%%
%%%%%%%%%%%%%%%%%%%%%%%%    Dirac Brackets     %%%%%%%%%%%%%%%%%%%%%%%%%%%%%%%%%%%%%%%%%%%%%%%%%%%%%%%%%%%%%
%%%%%%%%%%%%%%%%%%%%%%%%%%%%%%%%%%%%%%%%%%%%%%%%%%%%%%%%%%%%%%%%%%%%%%%%%%%%%%%%%%%%%%%%%%%%%%%%%%%%%%%%%%%%

\section{Dirac Brackets}\label{Section-Dirac_brackets}
\hs The first level of the Hamilton-Dirac approach to the dynamics has been completed in the previous section.
In the case when physical system possesses the second class constraints $\C_A$ standard Poisson bracket
can not be quantized by usual rule
\begin{eqnarray}
i\hbar\varrho\left(\{A,B\}\right)|\psi\rangle=\left[\varrho(A),\varrho(B)\right]|\psi\rangle,
\nonumber
\end{eqnarray}
where $\varrho$ is a representation of basic variables, since in the case when $A$, $B$ are 
the constraints $C_A$ then there is zero vector $\left(\varrho(C_A)\varrho(C_B)-
\varrho(C_B)\varrho(C_A)\right)|\psi\rangle$
on the right-hand side while the operator on the left-hand side $\varrho\left(\{C_A,C_B\}\right)$ is invertible. 
Hence there exists only one possibility for all
physical states solving quantum analogue of classical constraints represented by quantum equation
$\varrho(C_A)|\psi\rangle=0$ given by $|\psi\rangle=0$. Dirac solved this problem by introducing new
bracket and quantization is formulated by the representation of the Dirac instead of the Poisson algebra
(See details in \cite{Dirac}). Let $\C_A$ be the second class contraints and so $\{C_A,C_B\}=U_{AB}$ is 
invertible; then Dirac brackets are defined by
\begin{eqnarray}
\{A,B\}^*=\{A,B\}-\{A,C_A\}U^{AB}\{C_B,B\},
\end{eqnarray}
where $U_{AB}U^{BC}=\delta^C_A$. We divide our job in two parts. In the first part we define certain simple
brackets $\{\, ,\,\}'$  and then we use these partial brackets in the definition of the final Dirac brackets
 $\{\,,\,\}^*$.\\
\hs Let us define weak equivalence before we start our analysis of constraints. We say that two variables $A$, 
$A'$ are weakly equivalent, $A\hat{=}A'$, if their difference is proportional to the second class constraints. 
The second class constraints for our system are ($\bb^a$, $\BB^{ab}$, $M_{ab}$ are arbitrary)
\begin{eqnarray}
\!\!&\pp(\bb)&=\int\limits_{\Sigma}\pp_a\wedge\bb^a\nonumber=
\int\limits_{\Sigma}\tilde{p}^{\alpha}_a b^a_{\alpha}\dd^3x,\\
\!\!&\PP(\BB)&=\!\int\limits_{\Sigma}\frac{1}{2}\left(\pp_{ab}+\frac{1}{16\pi\kappa}\bleps_{abcd}
\EE^c\wedge\EE^d\right)\!\!\wedge\BB^{ab}=
	\!\int\limits_{\Sigma}\frac{1}{2}\PP_{ab}\wedge\BB^{ab}=
	\!\int\limits_{\Sigma}\frac{1}{2}\tilde{P}^{\alpha}_{ab}B_{\alpha}^{ab}\dd^3x,\nonumber\\
\!\!&\SSS(M)&=\int\limits_{\Sigma}\frac{1}{8\pi\kappa}M_{ab}\EE^{a}_{\phantom{c}}\EE^{b}_c\wedge
\D\EE^c
=\int\limits_{\Sigma}M_{ab}\SSS^{ab}=\int\limits_{\Sigma}M_{ab}\tilde{S}^{ab}\dd^3x.\nonumber
\end{eqnarray}
We start the analysis by their decompositions
\begin{eqnarray}
\ppkol=\pp_a\lambda^a\,\,\,\,\,\,\,\,\,\,\,\,\,\,\,\,\,\,\,\,\,\,\,\,\,\,\,\,\,\,\,\,\,\,
			\,\,\,\longleftrightarrow\,\,\,
	&&\pkol{\alpha}=\tilde{p}^{\alpha}_a\lambda^a,\nonumber\\
\pprov_a=\EE^{\bar{a}}_a\pp_{\bar{a}}\,\,\,\,\,\,\,\,\,\,\,\,\,\,\,\,\,\,\,\,\,\,\,\,\,\,\,\,\,\,\,\,\,\,
			\,\,\,\longleftrightarrow\,\,\,
	&&\prov{\alpha}{a}=\EE^{\bar{a}}_a\tilde{p}^{\alpha}_{\bar{a}},\nonumber\\
\PPkol_a=\PP_{ab}\lambda^b\,\,\,\,\,\,\,\,\,\,\,\,\,\,\,\,\,\,\,\,\,\,\,\,\,\,\,\,\,\,\,\,
			\,\,\,\longleftrightarrow\,\,\,
	&&\Pkol{\alpha}{a}=\tilde{P}^{\alpha}_{ab}\lambda^b,\nonumber\\
\PProv_a=\EE^{\bar{a}}_a\PP_{\bar{a}b}\wedge\EE^b\,\,\,\,\,\,\,\,\,\,\,\,\,\,\,\,\,\,
			\,\,\,\longleftrightarrow\,\,\,
	&&\Prov{a}=\EE^{\bar{a}}_a\tilde{P}^{\alpha}_{\bar{a}b}E^b_{\alpha},\nonumber\\
\bsigma^{ab}=\PP_{\bar{a}\bar{b}}\lambda_{\bar{c}}\wedge\bar{\bleps}^{\bar{a}\bar{b}\bar{c}(a}\EE^{b)}\,\,\,
			\,\,\,\longleftrightarrow\,\,\,
	&&\tsigma^{ab}=\tilde{P}^{\alpha}_{\bar{a}\bar{b}}\lambda_{\bar{c}}
		\bar{\bleps}^{\bar{a}\bar{b}\bar{c}(a}E^{b)}_{\alpha}.\nonumber
\end{eqnarray}
\hs Now we are going to eliminate constraints $\ppkol$, $\pprov_a$ and their "canonical friends" $\PPkol_{a}$, 
$\PProv_a$ by introducing "partial Dirac bracket" $\{\,,\,\}'$. This bracket plays important role even in 
the context of full Dirac bracket. In order to introduce it we need following expressions
\begin{eqnarray}
\{\PP(\BB),\pp(\bb)\}&=&\ints\frac{1}{16\pi\kappa}\bleps_{abcd}\BB^{ab}\wedge\bb^c\wedge\EE^d\nonumber\\
&\updownarrow&\nonumber\\
\{\tilde{P}^{\alpha}_{ab}(\xxx),\tilde{p}^{\beta}_c(\y)\}&=&\frac{1}{8\pi\kappa}\nonumber
\bleps_{abcd}\bar{\varepsilon}^{\alpha\beta\gamma}E^d_{\gamma}\delta_{\xxx\y}.
\end{eqnarray}
Hence nontrivial Poisson brackets are
\begin{eqnarray}
\{\PP(\BB^{^{\bot}}),\pp(\bb^{^{||}})\}&\hat{=}&
\ints-\frac{1}{8\pi\kappa}\bleps_{abcd}\BB^{a}\wedge\bb^b\lambda^c\wedge\EE^d\nonumber\\
&\updownarrow&\nonumber\\
\{\Pkol{\alpha}{a}(\xxx),\prov{\beta}{b}(\y)\}&\hat{=}&-\frac{1}{8\pi\kappa}
\bleps_{abcd}\bar{\varepsilon}^{\alpha\beta\gamma}\lambda^cE^d_{\gamma}\delta_{\xxx\y}=
		U^{\alpha\beta}_{ab}\delta_{\xxx\y},\nonumber\\
\,\nonumber\\\,\nonumber\\
\{\PP(\BB^{^{||}}),\pp(\bb^{^{\bot}})\}&\hat{=}&
\ints\frac{1}{8\pi\kappa}\bleps_{abcd}\B^{a}\bb\wedge\lambda^b\EE^c\wedge\EE^d\nonumber\\
&\updownarrow&\nonumber\\
\{\Prov{a}(\xxx),\pkol{\alpha}(\y)\}&\hat{=}&\frac{1}{8\pi\kappa}
\bleps_{abcd}\bar{\varepsilon}^{\alpha\beta\gamma}\lambda^bE^c_{\beta}E^d_{\gamma}\delta_{\xxx\y}=
		U^{\alpha}_{a}\delta_{\xxx\y}.\nonumber
\end{eqnarray}
It is easy to find that matrix $U^a_{\alpha}$ inverse to $U_a^{\alpha}$ is 
\begin{eqnarray}
U^a_{\alpha}=-\frac{4\pi\kappa}{e}E^a_{\alpha},\text{\hs where \hs }U^a_{\alpha}U^{\alpha}_b=\EE^a_b
\text{\hs and \hs}U^a_{\alpha}U^{\beta}_a=\delta^{\beta}_{\alpha}.
\end{eqnarray}
Next step is to look for the inverse matrix to $U_{ab}^{\alpha\beta}$. We can use ansatz 
$U^{ab}_{\alpha\beta}=AE^a_{\alpha}E^b_{\beta}+BE^a_{\beta}E^b_{\alpha}$ and the result is given by 
the expression
\begin{eqnarray}
U^{ab}_{\alpha\beta}=-\frac{4\pi\kappa}{e}(E^a_{\alpha}E^b_{\beta}-2E^a_{\beta}E^b_{\alpha}),
\text{\hs where \hs} U^{ab}_{\alpha\beta}U_{bc}^{\beta\gamma}=\EE^a_c\delta^{\gamma}_{\alpha}.
\end{eqnarray}
Now we have prepared everything what we need in order to define the partial Dirac bracket as follows
\begin{eqnarray}
\{A,B\}'=\{A,B\}&+&\ints\dd^3x\{A,\Pkol{\alpha}{a}(\xxx)\}U^{ab}_{\alpha\beta}(\xxx)\{\prov{\beta}{b}(\xxx),B\}\nonumber\\
		&-&\ints\dd^3x\{B,\Pkol{\alpha}{a}(\xxx)\}U^{ab}_{\alpha\beta}(\xxx)\{\prov{\beta}{b}(\xxx),A\}\nonumber\\
&+&\ints\dd^3x\{A,\Prov{a}(\xxx)\}U^a_{\alpha}(\xxx)\{\pkol{\alpha}(\xxx),B\}\nonumber\\
&-&\ints\dd^3x\{B,\Prov{a}(\xxx)\}U^a_{\alpha}(\xxx)\{\pkol{\alpha}(\xxx),A\}.\nonumber
\end{eqnarray}
The final Dirac bracket is going to be introduced within partial brackets and remaing constraints 
$\bsigma^{ab}$, $\SSS^{ab}$. First of all we should mention the following property of the partial bracket. Let
$A$ be an arbitrary variable on full phase space; then
\begin{eqnarray}
\{\bsigma(m),A\}'\hat{=}\{\bsigma(m),A\}, \nonumber
\end{eqnarray}
since 
\begin{eqnarray}
\{\bsigma(m),\pp(\bb)\}=\{\PP(\BB^m),\pp(\bb)\}\hat{=}
\int-\frac{1}{4\pi\kappa}\delta^{\bar{a}\bar{b}}_{ab}\bb^a\lambda_{\bar{a}}m_{\bar{b}\bar{c}}
\wedge\EE^{\bar{c}}\wedge\EE^b=0\nonumber
\end{eqnarray} 
and we have also $\{\bsigma(m),\PP(\BB)\}=\{\PP(\BB^m),\PP(\BB)\}=0$.
Hence we have as a consequence
\begin{eqnarray}
\{\bsigma(m),\bsigma(m')\}'\hat{=}0 \,\,\,\,\longleftrightarrow\,\,\,\,
\{\tsigma^{ab}(\xxx),\tsigma^{cd}(\y)\}\hat{=}0.\nonumber
\end{eqnarray}
Next important classical commutator is 
\begin{eqnarray}
\{\bsigma(m),\SSS(M)\}'\hat{=}\{\bsigma(m),\SSS(M)\}=\ints-\frac{1}{8\pi\kappa}m_{a\bar{a}}M_{b\bar{b}}
\eta_{c\bar{c}}\lambda_d\lambda_{\bar{d}}\bar{\bleps}^{abcd}\bar{\bleps}^{\bar{a}\bar{b}\bar{c}\bar{d}}\blome,
\end{eqnarray}
where $\blome=\frac{1}{3!}\bleps_{abcd}\lambda^a\EE^b\wedge\EE^c\wedge\EE^d=e\,\dd^3x$. Now it is time to pay
debt from the previous section where we have stated that $\tilde{G}^{a\bar{a}b\bar{b}}$ is invertible. We are 
going to do even more. We are going to calculate inverse of $U^{a\bar{a}b\bar{b}}=\frac{e}{8\pi\kappa}
\tilde{G}^{a\bar{a}b\bar{b}}$.
We can write
\newcommand{\Sti}{\tilde{S}}
\begin{eqnarray}
\{\bsigma(m),\SSS(M)\}&=&\ints \frac{1}{8\pi\kappa}m_{a\bar{a}}M_{b\bar{b}}
\tilde{G}^{a\bar{a}b\bar{b}}\blome
\nonumber\\
&\updownarrow&\nonumber\\
\{\tsigma^{a\bar{a}}(\xxx),\Sti^{b\bar{b}}(\y)\}&=&\frac{e}{8\pi\kappa}\tilde{G}^{a\bar{a}b\bar{b}}
\delta_{\xxx\y}=U^{a\bar{a}b\bar{b}}\delta_{\xxx\y}
\nonumber
\end{eqnarray}
and
\begin{eqnarray}
\tilde{G}^{a\bar{a}b\bar{b}}=-\frac{1}{2}\eta_{c\bar{c}}\lambda_d\lambda_{\bar{d}}
\left(\bar{\bleps}^{abcd}\bar{\bleps}^{\bar{a}\bar{b}\bar{c}\bar{d}}+
		\bar{\bleps}^{\bar{a}bcd}\bar{\bleps}^{a\bar{b}\bar{c}\bar{d}}
\right).
\end{eqnarray}
Let us transform $U^{a\bar{a}b\bar{b}}$ into more suitable form. In order to do so we need to use
the spatial metric tensor which is due to our choice of signature strictly negative
\begin{eqnarray}
\q=\eta_{ab}\EE^a\otimes\EE^b=\q_{ab}\EE^a\otimes\EE^b=q_{\alpha\beta}\dd x^{\alpha}\otimes\dd x^{\beta},
\end{eqnarray}
where $\q_{ab}=\eta_{\bar{a}\bar{b}}\EE^{\bar{a}}_a\EE^{\bar{b}}_b$, its inverse matrix is $q^{\alpha\beta}q_{\beta\gamma}=\delta^{\alpha}_{\beta}$ or $\q^{ab}\q_{bc}=\EE^a_c$ and determinant
\begin{eqnarray}
q\varepsilon_{\alpha\beta\gamma}=q_{\alpha\bar{\alpha}}q_{\beta\bar{\beta}}q_{\gamma\bar{\gamma}}
\bar{\varepsilon}^{\bar{\alpha}\bar{\beta}\bar{\gamma}}.\nonumber
\end{eqnarray}
It should be noted that $\q^{ab}\neq\EE^a_{\bar{a}}\EE^b_{\bar{b}}\eta^{\bar{a}\bar{b}}$. Now we can write
\newcommand{\las}{{\lambda^{\!^*}}}
\begin{eqnarray}
U^{a\bar{a}b\bar{b}}=\frac{\,\,\,\,\,\las^2}{16\pi\kappa}(2\q^{a\bar{a}}\q^{b\bar{b}}-
\q^{ab}\q^{\bar{a}\bar{b}}-\q^{a\bar{b}}\q^{\bar{a}b}),
\end{eqnarray}
where we have used formula $q=-e^2\las^2$ and $\las^2=\eta^{ab}\lambda_a\lambda_b$. Now we are looking for 
inverse matrix to $U^{a\bar{a}b\bar{b}}$ in the form $U_{a\bar{a}b\bar{b}}=A\q_{a\bar{a}}\q_{b\bar{b}}+
B(\q_{ab}\q_{\bar{a}\bar{b}}+\q_{a\bar{b}}\q_{\bar{a}b})$ and the result is given by the expression
\begin{eqnarray}
U_{a\bar{a}b\bar{b}}=\frac{4\pi\kappa}{\las^2\,e}(\q_{a\bar{a}}\q_{b\bar{b}}
-\q_{ab}\q_{\bar{a}\bar{b}}-\q_{a\bar{b}}\q_{\bar{a}b}),\,\,\,\text{where}\,\,\,
U^{a\bar{a}b\bar{b}}U_{b\bar{b}c\bar{c}}=\EE^{(a}_{\phantom{(}c}\EE^{\bar{a})}_{\bar{c}\phantom{)}}.
\end{eqnarray}
Finally we can define the full Dirac bracket as
\begin{eqnarray}
\{A,B\}^*&=&\{A,B\}'+
\int\!\!\dd^3x\{A,\tsigma^{a\bar{a}}(\xxx)\}'U_{a\bar{a}b\bar{b}}(\xxx)
	\{\Sti^{b\bar{b}}(\xxx),B\}'-\nonumber\\
&&\,\,\,\,\,\,\,\,\,\,\,\,\,\,\,\,\,\,\,\,
-\int\!\!\dd^3x\{B,\tsigma^{a\bar{a}}(\xxx)\}'U_{a\bar{a}b\bar{b}}(\xxx)
	\{\Sti^{b\bar{b}}(\xxx),A\}'-
\nonumber\\
&&-\!\!\int\!\!\dd^3x\dd^3y\{A,\tsigma^{a\bar{a}}(\xxx)\}'U_{a\bar{a}b\bar{b}}(\xxx)
\{\Sti^{b\bar{b}}(\xxx),\Sti^{c\bar{c}}(\y)\}'U_{c\bar{c}d\bar{d}}(\y)\{\tsigma^{d\bar{d}}(\y),B\}'.\nonumber
\end{eqnarray}
\hs In order to finish the phase space reduction we need to describe a reduced manifold. Let us start
with full phase space $\tilde{\Gamma}$ described by canonical variables $\lambda^a$, $\blpi_a$, $\dots$, 
$\U_{ab}$ (see table \ref{tabulka}). As we have seen in section \ref{section-Hamiltonian} the first reduction
is given by $\C^{ab}=\CC^{ab}=\blPHI_{ab}=\U_{ab}=0$ while conditions $\blphi=\uu=0$ are the first class 
contraints. These contraints mean that $\B$, $\BB$ are arbitrary and physics does not depend on them.
Hence we can write $\tilde{\Gamma}\vb{\text{red}}=\hat{\Gamma}\times\Lambda\Sigma$, where $\Lambda\Sigma$
is Cartan algebra of all forms on $\Sigma$ of variables $\B$, $\dots$, $\uu$ and $\hat{\Gamma}$ is described by variables $\lambda^a$, $\dots$, $\pp_{ab}$. Whole dynamics takes place in $\hat{\Gamma}$. Let us consider a set
\begin{eqnarray}
\Conf=\{(\lambda^a,\EE^a);\, e>0,\, \eta_{ab}\lambda^a\lambda^b>0,\, \lambda^0>0, \, 
\q<0\}.\nonumber
\end{eqnarray}
Hence due to condition $e>0$ we have $\Conf\subset\GL^+(\Mkw)$. However $\Conf$ is not a group.
\nobreak{Nevertheless} for every sufficiently small change $(\Delta\lambda^a,\Delta\EE^a)$ the new element 
is again from $\Conf$, i.e. $(\lambda^a+\Delta\lambda^a,\EE^a+\Delta\EE^a)\in\Conf$; in other words 
$\Conf$ is a manifold. Hence we can construct canonically its cotangent bundle $\Td^*\Conf=\Td_1\Conf$ with 
symplectic structure $\omega_{\Conf}$ on it. $\Td^*\Conf$ is described by canonical coordinates
$(\lambda^a$, $\EE^a$, $\blpi_a$, $\pp_a)$. Another structure of $\hat{\Gamma}$ is given by space
\begin{eqnarray}
\mathfrak{G}=(\Lambda_0\Ad^2\Sigma\times\Lambda_3\Ad_2\Sigma)\times		
				 (\Lambda_1\Ad^2\Sigma\times\Lambda_2\Ad_2\Sigma)
\end{eqnarray}
described by variables $(\Lambda^{ab}$, $\blPI_{ab}$; $\AAA^{ab}$, $\pp_{ab})$. Hence 
$\hat{\Gamma}=\Td^*\Conf\times\mathfrak{G}$.\\
\hs Since $\AAA^{ab}$ is antisymmetric matrix 1-form we can decompose it as
\begin{eqnarray}
\AAA^{ab}=2\AAA^{[a}\lambda^{b]}+2\A^{\bar{a}}\EE^{[a}_{\,\bar{a}}\EE^{b]}_{\phantom{\bar{a}}}+
			 \bar{\bleps}^{ab\bar{a}\bar{b}}\EE^{\bar{c}}\lambda_{\bar{a}}\alpha_{\bar{b}\bar{c}},
\end{eqnarray}
where $\alpha_{ab}=\alpha_{ba}$. Relevant information about $\AAA^a$ and $\A^a$ is encoded in a new variable
\newcommand{\QQ}{\mathbf{Q}}
\begin{eqnarray}
\FFF_{a}=\frac{1}{2}\bleps_{abcd}\AAA^{bc}\wedge\EE^d\,\,\,\,\longleftrightarrow\,\,\,\,
\FFF(\KK)=\ints\frac{1}{2}\bleps_{abcd}\KK^a\wedge\AAA^{bc}\wedge\EE^d,
\end{eqnarray}
while $\alpha_{ab}$ does not enter $\FFF_a$. Since $\{\bsigma(m),\FFF(\KK)\}'\hat{=}0$ and 
$\{\bsigma(m),\EE(\QQ)\}'\hat{=}0$, where $\EE(\QQ)=\ints\QQ_a\wedge\EE^a$ we have that
\begin{eqnarray}
\begin{matrix}
\{\,\EE(\QQ),\FFF(\KK)\,\}^*&\hat{=}&\{\,\EE(\QQ),\FFF(\KK)\,\}'&=&-8\pi\kappa\ints\QQ_a\wedge\KK^a,\\ 
\{\FFF(\KK),\FFF(\KK ')\}^*&\hat{=}&\{\FFF(\KK),\FFF(\KK ')\}'&=&0.\,\,\,\,\,\,\,\,\,\,\,\,\,\,\,\,\,\,\,\,\,\,\,\,\,\,\,\,\,\,\,\,\,\,\,\,\,\,\,
\end{matrix}
\label{DB-EF}
\end{eqnarray}
Analogously, we obtain the rest of Dirac brackets for our variables on $\hat{\Gamma}$. The nontrivial 
results are
\begin{eqnarray}
%\begin{matrix}
%&\{\, \lambda^a\,,\,\blpi(\mu)\,\}^*&\hat{=}&\mu^a,\\
%&\{\Lambda^{ab},\blPI(\Gamma)\}^*&\hat{=}&\Gamma^{ab}.
%\end{matrix}
\{\, \lambda^a\,,\,\blpi(\mu)\,\}^*&\hat{=}&\mu^a,\label{DB-lamdapi}\\
\{\Lambda^{ab},\blPI(\Gamma)\}^*&\hat{=}&\Gamma^{ab}.
\end{eqnarray}
The reduction of $\hat{\Gamma}$ is almost finished. We can express $\alpha_{ab}$ from the condition 
$\SSS^{ab}=0$ as function(al) of $\lambda^a$, $\EE^a$ and $\FFF_a$. The remaining second class contraints 
are trivially soluble. Since variables $\B$, $\dots$, $\uu$ do not describe any dynamics we can 
cast them away by additional fixation $\B=0$ and $\BB=0$. Similar, we can proceed with $\Lambda^{ab}$. Hence 
we have the final reduced phase space 
\begin{eqnarray}
\Gamma=\Td^*\Conf
\end{eqnarray}
described by variables $(\lambda^a$, $\EE^a$, $\blpi_a$, $\FFF_a)$ with symplectic structure defined 
by (\ref{DB-EF}) and (\ref{DB-lamdapi}).

%%%%%%%%%%%%%%%%%%%%%%%%%%%%%%%%%%%%%%%%%%%%%%%%%%%%%%%%%%%%%%%%%%%%%%%%%%%%%%%%%%%%%%%%%%%%%%%%%%%%%%%%%%%
%%%%%%%%%%%%%%%%%%%%%%%%%%%%   Discussion   %%%%%%%%%%%%%%%%%%%%%%%%%%%%%%%%%%%%%%%%%%%%%%%%%%%%%%%%%%%%%%%
%%%%%%%%%%%%%%%%%%%%%%%%%%%%%%%%%%%%%%%%%%%%%%%%%%%%%%%%%%%%%%%%%%%%%%%%%%%%%%%%%%%%%%%%%%%%%%%%%%%%%%%%%%%

\section{Discussion and Open Problems}\label{Section-Discussion_and_Open_Problems}
\hs In section \ref{Section-Lagrangian_of_Cartan_Theory} we have started with the 
orthonormal coframe $\ee^a$ and the general gravitational connection $\hat{\nabla}$ described by its 
forms $\GGamma^{ab}=\eta^{b\bar{b}}\GGamma^a_{\phantom{a}\bar{b}}$. We have derived equations of motion 
which have fixed $\GGamma^{ab}=\HAA^{ab}+\HBB\eta^{ab}$ where $\HAA^{ab}$ is related to the metric connection 
$\HD$ and $\HBB$ is an arbitrary 1-form. The torsion of $\HD$ vanishes as a consequence of EOM, hence 
$\HAA^{ab}$ can be expressed as a functional of coframe $\ee^a$ which is given by the solution of Einstein 
equations. We have induced the geometrical 
structure on the spatial section $\Sigma$ inherited from the spacetime $\M$ and hence $\SO(\g)$ is still 
(part of) the gauge freedom which is in opposite to the standard euclidean loop formulation of gravity where the 
orthonormal coframe $\ee^a$ is fixed to be tangential to $\Sigma$ in spatial covectors and its time covector 
is normal to $\Sigma$. Then we have used $\SO(\g)$ structure in the Hamilton-Dirac formulation of the 
Einstein-Cartan theory. Since our system is degenerated and it contains both classes of constraints the 
Dirac bracket has been introduced. The Dirac procedure has been finished by introducing the reduced phase 
space described by coordinates $(\lambda^a,\EE^a, \blpi_a, \FFF_a)$.\\
\hs The first class constraints have to be analyzed. If $\Sigma$ is noncompact manifold 
then the first class constraints $\RR(\mu)$, $\TT(\Lambda)$ does not belong to the set of 
bounded function(al)s over the phase space if meaningful fall off conditions are suggested. Thus surface
terms should be added in a similar way as in ADM formulation of General Relativity and 
the final Hamiltonian is not identically vanishing in this case. The analysis of the first class
constraints including their algebraic properties will be performed in next paper of this series.\\
\hs Another problem which should be explored in detail is the existence of the second class constraints 
$\bsigma(m)$ and $\SSS(M)$. We already know how to deal with this on the classical level. But since the goal is a quantization of the Einstein-Cartan theory then they may cause insurmountable obstacle of the 
whole theory. There are few ways how to solve such kind of problems. The first thing which can be done
is to solve the constraints classically as we have indicated in section \ref{Section-Dirac_brackets} and 
then quantize the rest of variables.
Another possibility is inspired by known property in quantum mechanics. 
Let $C$ and $K$ be the second class constraints with Poisson bracket $\{C,K\}=1$. Then one can construct 
its representation $\varrho$ on an appropriate Hilbert space and proceeds to construct the
creation and annihilation operator $\hat{a}^{+/-}$ associated with them. 
If one finds vacuum of these operators, i.e. $\hat{a}^-|\Omega\rangle = 0$  then 
$\langle\Omega|C|\Omega\rangle=\langle\Omega|K|\Omega\rangle=0$ and these constraints are solved at least 
for mean values. The third idea how to solve these constraints lies in enlargement of the phase space in such 
a way, that these constraints become first class. We know from observation of 2+1 Einstein-Cartan theory 
that constraints $\TT_{ab}$ are good candidates for generators of Lorentz algebra. Constraints $\SSS^{ab}$
look similar. Both of them look like $\EE^a\wedge\D\EE^b$. Constraints $\TT_{ab}$ are antisymmetric in $ab$
while $\SSS^{ab}$ are symmetric. Hence it looks like something which should match together and question is
what happens if we enlarge phase space by metric variable $g_{ab}$, i.e. if we broke gauge fixing represented
by orthonormality of frames. Since $\SSS^{ab}$ lives on spatial part of $\Td\Mkw$ if this idea works then 
the resulting group cannot be whole $\GL(\Mkw)$. Spatial metric has six degrees of freedom plus six for 
its momentum is twelve which is exactly the number of degrees of freedom fixed by constraints 
$\SSS^{ab}$ and $\bsigma^{ab}$. 
Anyway if this would work then this idea can not be used in higher dimensional space time. We can still 
use ghosts and quite general theory behind them.

\begin{acknowledgments}
I would like to thank to prof. Ji\v r\' i Bi\v c\' ak for his patience, support 
in hard times, neverending discussions, etc... My thanks also belong to Otakar Sv\' itek for practically 
the same reasons, to Vladim\' ir Balek for opening closed eyes at the beginning of this work,
Mari\' an Fecko for his excellent book giving many inspirations, Michal Demetrian, whole "old" ITP
in Prague and also Department of Theoretical physics in Bratislava, to Jakub Lehotsk\' y, V\' aclav D\v edi\v c, Richard Richter, Zden\v ek Soukup for their friendship. I am very appreciating the support, love of and life experience with my Femme Fatale Hana Korbov\' a. Unfortunately I am no more able to thank to my mum thus please let me dedicate this work to her, to \v Ludmila Pilcov\' a in memoriam.
\end{acknowledgments}

%%%%%%%%%%%%%%%%%%%%%%%%%%%%%%%%%%%%%%%%%%%%%%%%%%%%%%%%%%%%%%%%%%%%%%%%%%%%%%%%%%%%%%%%%%%%%%%%%%%%%%%%%%%%%
%%%%%%%%%%%%%%%%%%%%%%%%%%%%%%%%%%%%%   NOTATION AND CONVENTIONS    %%%%%%%%%%%%%%%%%%%%%%%%%%%%%%%%%%%%%%%%%
%%%%%%%%%%%%%%%%%%%%%%%%%%%%%%%%%%%%%%%%%%%%%%%%%%%%%%%%%%%%%%%%%%%%%%%%%%%%%%%%%%%%%%%%%%%%%%%%%%%%%%%%%%%%%
\appendix
\section{Notation and conventions}\label{notation}
$\phantom{.}$\\
Manifold structure and indices:\\
$\M$ - spacetime, $\Sigma$ - spatial section of $\M=\R\times\Sigma$\\
$\Sigma$ - spatial section in $\M$\\
$\Mkw$ - tangential Minkowski space (see section \ref{Section-3+1_Decomposition})\\
$a,b,\dots=0,1,2,3$ - frame indices\\
$\mu,\nu,\dots=0,1,2,3$ - spacetime coordinate indices\\
$\alpha,\beta,\dots=1,2,3$ - spatial coordinate indices\\
$\eta_{ab}$ - Minkowski metric with signature $(+,-,-,-)$\\\\
Groups:\\
$\GL(V)$ - general linear group over (real) vector space $V$\\
$\GL^+(V)$ - positive general linear group over (real) vector space $V$(elements of $\GL(V)$ with 
positive determinant)\\
$\OOO(\g)$ - orthonormal group over metric vector space $(V,\g)$ or manifold $(\M,\g)$\\
$\overline{\SO}(\g)\subset\OOO(\g)$ - special orthonormal group over vector space $(V,\g)$ or manifold $(\M,\g)$\\
$\SO(\g)\subset\overline{\SO}(\g)$ - proper Lorentz group over vector space $(V,\g)$ or manifold $(\M,\g)$ preserving 
righthand and future time orientation\\
\\
(Anti)symmetrization:\\\\
$A^{[ab]}=\frac{1}{2}(A^{ab}-A^{ba})$\\
$S^{(ab)}=\frac{1}{2}(S^{ab}+S^{ba})$\\
etc.\\\\
Antisymmetric delta and Levi-Civita symbol:\\\\
$\delta^{a\dots b}_{c\dots d}=\delta^{[a}_{\phantom{[}c}\dots \delta^{b]}_{d\phantom{]}}=
\delta_{[c}^{\phantom{[}a}\dots \delta_{d]}^{b\phantom{]}}=
\delta^{[a}_{[c}\dots \delta^{b]}_{d]}$\\
$\bleps_{abcd}=\bleps_{[abcd]}$, $\bar{\bleps}^{abcd}=\bar{\bleps}^{[abcd]}$ and 
$\bleps_{0123}=\bar{\bleps}^{0123}=1$\\
$\varepsilon_{\alpha\beta\gamma}=\varepsilon_{[\alpha\beta\gamma]}$, 
$\bar{\varepsilon}^{\alpha\beta\gamma}=\bar{\varepsilon}^{[\alpha\beta\gamma]}$ and 
$\varepsilon_{123}=\bar{\varepsilon}^{123}=1$\\
\\
Tensor spaces:\\
$(\Td\M,\otimes)$ - algebra of all tensors over the spacetime manifold $\M$\\
$\Td^p_q\M$ - (real) vector space of all tensors of rank $^p_q$ over the spacetime manifold $\M$\\
$(\Td\Mkw,\otimes)$ - algebra of all tensors in $\Mkw$\\
$\Td^p_q\Mkw$ - (real) vector space of all tensors of rank $^p_q$ in $\Mkw$\\
\\
Cartan algebra and exterior product:\\
$(\Lambda\M,\wedge)$ - Cartan algebra of all spacetime forms. $\Lambda_p\M$ - space of spacetime p-forms.\\
$(\Lambda\Sigma,\wedge)$ - Cartan algebra of all spatial forms. $\Lambda_p\Sigma$ - 
space of spatial p-forms.\\
$(\Lambda\Td\M,\wedge)$ - algebra of forms with values in tensor space $\Td\M$\\
$\Lambda\Td^p_q\M$ - (real) vector space of forms with values in tensor space $\Td^p_q\M$\\
$(\Lambda\Td\Mkw,\wedge)$ - algebra of forms with values in tensor space $\Td\Mkw$\\
$\Lambda\Td^p_q\Mkw$ - (real) vector space of forms with values in tensor space $\Td^p_q\Mkw$\\
If $\alpha, \beta \in \Lambda_1\M\text{ or }\Lambda_1\Sigma$ then $\alpha\wedge\beta=
\alpha\otimes\beta-\beta\otimes\alpha$\\
$\dd^4 x=\dd t\wedge\dd^3x$, $\dd^3 x=\dd x^1\wedge\dd x^2\wedge\dd x^3$\\
\\
Interior product:\\
$(i_v\alpha)(u_1,...,u_{p-1})=\alpha(v,u_1,...,u_{p-1})$ $\forall \alpha\in\Lambda_p\M$ 
or $\Lambda_p\Sigma$\\
\\
Derivative operators:\\
$\hd$ - exterior derivative operator on spacetime $\Lambda\M$. Anyway we write $\dd t=\hd t$\\
$\dd$ - spatial exterior derivative operator on $\Lambda\Sigma$\\
$\hat{\nabla}$ - general covariant exterior derivative operator on $\Lambda\Td\M$, or general connection 
associated with $\GGamma^a_{\phantom{a}b}$\\
$\HD$ - $\SO(\g)$-covariant exterior derivative operator on $\Lambda\Td\M$ associated with 
$\HAA^a_{\phantom{a}b}=\eta_{bc}\HAA^{ac}$\\
$\D$ - spatial $\SO(\g)$-covariant exterior derivative operator on $\Lambda\Td\Mkw$ associated with $\AAA^a_{\phantom{a}b}=\eta_{bc}\AAA^{ac}$

\section{2+1 Dimensional  Einstein-Cartan Theory}\label{2+1GR}
If we already start with metric-compatible connection (Similar analysis of general connection can be done as in 3+1 case, but for simplicity we fix connection to be metric-compatible already now.) then 
Lagrangian for 2+1 dimensional Einstein-Cartan theory can be written as
\begin{eqnarray}
\LL=\frac{1}{2}\bleps_{abc}\HR^{ab}\wedge\ee^c.
\end{eqnarray}
EOM:
\begin{eqnarray}
\HR^{ab}&=&0,\\
\hat{\TT}^a\,\,&=&0.
\end{eqnarray}
Using 2+1 decomposition 
\begin{eqnarray}
\ee^a\,\,&=&\lambda^a\dd t+\EE^a,\nonumber\\
\HAA^{ab}&=&\Lambda^{ab}\dd t+\AAA^{ab}\nonumber
\end{eqnarray}
leads to Hamiltonian:
\begin{eqnarray}
\hamcel=\blpi(\nu)+\blPI(\Gamma)+\pp(\bb)+\PP(\BB)+\RR(\lambda)+\TT(\Lambda),
\end{eqnarray}
where
\begin{eqnarray}
&\blpi(\nu)&=\ints\nu^a\wedge\blpi_a,\\
&\blPI(\Gamma)&=\ints\frac{1}{2}\Gamma^{ab}\wedge\blPI_{ab},\\
&\pp(\bb)&=\ints\bb^a\wedge\pp_a,\\
&\PP(\BB)&=\ints\frac{1}{2}\BB^{ab}\wedge(\pp_{ab}-\bleps_{abc}\EE^c),\\
&\RR(\lambda)&=\ints-\frac{1}{2}\bleps_{abc}\lambda^{a}\HR^{bc},\\
&\TT(\Lambda)&=\ints-\frac{1}{2}\bleps_{abc}\Lambda^{ab}\D\EE^{c}.
\end{eqnarray}
Momenta and velocities variables are given by table \ref{tabulka_appendix}.\\
Primary constraints are
\begin{eqnarray}
&\blpi(\nu)&= 0,\nonumber\\
&\pp(\bb)&= 0,\nonumber\\
&\blPI(\Gamma)&= 0,\nonumber\\
&\PP(\BB)&= 0. \nonumber
\end{eqnarray}
Poisson brackets between Hamiltonian and $\pp(\bb)$ or $\PP(\BB)$ lead to Lagrange multipleirs
\begin{table}[t]
\caption{Table of basic variables}\label{tabulka_appendix}
\begin{center}
    \begin{tabular}{| l | l | l |}
    \hline
    Variables & Momentum & Velocities \\ \hline
    $\lambda^a$ 
	& 
	 $\blpi_a=\tilde{\pi}_a\dd^2 x$ where $\tilde{\pi}_a=\partial \Lg / \partial\dot{\lambda}^a$ 
	& 
	 $\nu^a=\dot{\lambda}^a$
   \\ 
    $\EE^a=E^a_{\alpha}\dd x^{\alpha} $  
	&
	 $\pp_a=\tilde{p}_a^{\alpha}\varepsilon_{\alpha\beta}\dd x^{\beta}$ where 
		$\tilde{p}_a^{\alpha}=\partial \Lg / \partial\dot{E}^a_{\alpha}$ 
	& 
		$\bb^a=\dot{\EE}^a$
   \\ 
    $\Lambda^{ab}$ 
	& 
	 $\blPI_{ab}=\tilde{\Pi}_{ab}\dd^2 x$ where $\tilde{\Pi}_{ab}=\partial \Lg / \partial\dot{\Lambda}^{ab}$
	& 
	 $\Gamma^{ab}=\dot{\Lambda}^{ab}$
	\\ 
    $\AAA^{ab}=A^{ab}_{\alpha}\dd x^{\alpha}$ 
	& 
	 $\pp_{ab}=\tilde{p}_{ab}^{\alpha}\varepsilon_{\alpha\beta}\dd x^{\beta}$ where
	 $\tilde{p}_{ab}^{\alpha}=\partial \Lg / \partial\dot{A}^{ab}_{\alpha}$ 
	& 
	 $\BB^{ab}=\dot{\AAA}^{ab}$ 
	\\ \hline
    \end{tabular}
\end{center}
\end{table}
\begin{eqnarray}
\BB^{ab}&=&\D\Lambda^{ab},\nonumber\\
\bb^a\,\,&=&\D\lambda^a-\eta_{\bar{a}\bar{b}}\Lambda^{a\bar{a}}\EE^{\bar{b}},\nonumber
\end{eqnarray}
while $\blpi(\nu)$ and $\blPI(\Gamma)$ give new constraints
\begin{eqnarray}
&\RR(\nu)&=\int\limits_{\Sigma}-\frac{1}{2}\bleps_{abc}\nu^{a}\HR^{bc},\nonumber\\
&\TT(\Gamma)&=\int\limits_{\Sigma}-\frac{1}{2}\bleps_{abc}\Gamma^{ab}\D\EE^{c}.\nonumber
\end{eqnarray}
No other new constraints appear and $\pp$, $\PP$ are the second class constrains. Next step is the 
definition of Dirac bracket thus we need to evaluate
\begin{eqnarray}
\left\{\PP(\tilde\BB),\pp(\tilde{\bb})\right\}=
\int\limits_{\Sigma}-\frac{1}{2}\bleps_{abc}\tilde{\BB}^{ab}\wedge\tilde{\bb}^{c},\nonumber
\end{eqnarray}
what is equal to
\begin{eqnarray}
\left\{\tilde{P}^{\alpha}_{ab}(x),\tilde{p}^{\beta}_{c}(y)\right\}=
-\bleps_{abc}\bar{\varepsilon}^{\alpha\beta}\delta_{xy}.\nonumber
\end{eqnarray}
Dirac bracket is defined as
\begin{eqnarray}
\{A,B\}^*=\{A,B\}&+&\int\frac{\dd x}{2} 
\{A,\tilde{P}^{\alpha}_{ab}\}\bar{\bleps}^{abc}\varepsilon_{\alpha\beta}\{\tilde{p}^{\beta}_c,B\}\nonumber\\
					  &-&\int\frac{\dd x}{2} \{B,\tilde{P}^{\alpha}_{ab}\}\bar{\bleps}^{abc}\varepsilon_{\alpha\beta}\{\tilde{p}^{\beta}_c,A\}	
\end{eqnarray}
and constraints algebra is given by commutators
\begin{eqnarray}
&&\{\RR(\mu),\RR(\nu)\}^*=0,\\
&&\{\RR(\mu),\TT(\Lambda)\}^*=-\RR(\Lambda\eta\mu),\\
&&\{\TT(\Lambda),\TT(\Gamma)\}^*=\TT(\tilde{\Lambda}),
\end{eqnarray}
where $\tilde{\Lambda}^{ab}=2\delta^{ab}_{\bar{a}\bar{b}}\Lambda^{\bar{a}\bar{c}}\eta_{\bar{c}\bar{d}}
\Gamma^{\bar{d}\bar{b}}$ and $(\Lambda\eta\mu)^a=\Lambda^{ab}\eta_{bc}\mu^c$. We see that the constraints 
of 2+1 dimensional Einstein-Cartan theory generate Poincar\' e algebra.

\bibliography{ECTheoryI}% Produces the bibliography via BibTeX.

\begin{thebibliography}{17}%
\makeatletter
\providecommand \@ifxundefined [1]{%
 \@ifx{#1\undefined}
}%
\providecommand \@ifnum [1]{%
 \ifnum #1\expandafter \@firstoftwo
 \else \expandafter \@secondoftwo
 \fi
}%
\providecommand \@ifx [1]{%
 \ifx #1\expandafter \@firstoftwo
 \else \expandafter \@secondoftwo
 \fi
}%
\providecommand \natexlab [1]{#1}%
\providecommand \enquote  [1]{``#1''}%
\providecommand \bibnamefont  [1]{#1}%
\providecommand \bibfnamefont [1]{#1}%
\providecommand \citenamefont [1]{#1}%
\providecommand \href@noop [0]{\@secondoftwo}%
\providecommand \href [0]{\begingroup \@sanitize@url \@href}%
\providecommand \@href[1]{\@@startlink{#1}\@@href}%
\providecommand \@@href[1]{\endgroup#1\@@endlink}%
\providecommand \@sanitize@url [0]{\catcode `\\12\catcode `\$12\catcode
  `\&12\catcode `\#12\catcode `\^12\catcode `\_12\catcode `\%12\relax}%
\providecommand \@@startlink[1]{}%
\providecommand \@@endlink[0]{}%
\providecommand \url  [0]{\begingroup\@sanitize@url \@url }%
\providecommand \@url [1]{\endgroup\@href {#1}{\urlprefix }}%
\providecommand \urlprefix  [0]{URL }%
\providecommand \Eprint [0]{\href }%
\providecommand \doibase [0]{http://dx.doi.org/}%
\providecommand \selectlanguage [0]{\@gobble}%
\providecommand \bibinfo  [0]{\@secondoftwo}%
\providecommand \bibfield  [0]{\@secondoftwo}%
\providecommand \translation [1]{[#1]}%
\providecommand \BibitemOpen [0]{}%
\providecommand \bibitemStop [0]{}%
\providecommand \bibitemNoStop [0]{.\EOS\space}%
\providecommand \EOS [0]{\spacefactor3000\relax}%
\providecommand \BibitemShut  [1]{\csname bibitem#1\endcsname}%
\let\auto@bib@innerbib\@empty
%</preamble>
\bibitem [{\citenamefont {{J. E. Nelson, C. Teitelboim}}(1978)}]{RLC-electron}%
  \BibitemOpen
  \bibfield  {author} {\bibinfo {author} {\bibnamefont {{J. E. Nelson, C.
  Teitelboim}}},\ }\href@noop {} {\bibfield  {journal} {\bibinfo  {journal}
  {Ann.\ Phys.}\ }\textbf {\bibinfo {volume} {116}},\ \bibinfo {pages} {86}
  (\bibinfo {year} {1978})}\BibitemShut {NoStop}%
\bibitem [{\citenamefont {{T. W. B. Kibble}}(1961)}]{Kibble}%
  \BibitemOpen
  \bibfield  {author} {\bibinfo {author} {\bibnamefont {{T. W. B. Kibble}}},\
  }\href@noop {} {\bibfield  {journal} {\bibinfo  {journal} {Jour. \ Math. \
  Phys.}\ }\textbf {\bibinfo {volume} {2}},\ \bibinfo {pages} {212} (\bibinfo
  {year} {1961})}\BibitemShut {NoStop}%
\bibitem [{\citenamefont {{F. W. Hehl, P. von der Heyde, G. D.
  Kerlick}}(1976)}]{Hehl1}%
  \BibitemOpen
  \bibfield  {author} {\bibinfo {author} {\bibnamefont {{F. W. Hehl, P. von der
  Heyde, G. D. Kerlick}}},\ }\href@noop {} {\bibfield  {journal} {\bibinfo
  {journal} {Rev. \ Mod. \ Phys.}\ }\textbf {\bibinfo {volume} {48}},\ \bibinfo
  {pages} {393} (\bibinfo {year} {1976})}\BibitemShut {NoStop}%
\bibitem [{\citenamefont {{F. W. Hehl, J. D. McCrea, E. W. Mielke, Y.
  Ne'eman}}(1995)}]{Hehl2}%
  \BibitemOpen
  \bibfield  {author} {\bibinfo {author} {\bibnamefont {{F. W. Hehl, J. D.
  McCrea, E. W. Mielke, Y. Ne'eman}}},\ }\href@noop {} {\bibfield  {journal}
  {\bibinfo  {journal} {Phys. \ Rep.}\ }\textbf {\bibinfo {volume} {258}},\
  \bibinfo {pages} {1} (\bibinfo {year} {1995})}\BibitemShut {NoStop}%
\bibitem [{\citenamefont {{J. Bi\v c\' ak}}(1966)}]{Bicak_pole}%
  \BibitemOpen
  \bibfield  {author} {\bibinfo {author} {\bibnamefont {{J. Bi\v c\' ak}}},\
  }\href@noop {} {\bibfield  {journal} {\bibinfo  {journal} {Czechoslovak \ J.
  \ Phys. \ B}\ }\textbf {\bibinfo {volume} {16}},\ \bibinfo {pages} {95}
  (\bibinfo {year} {1966})}\BibitemShut {NoStop}%
\bibitem [{\citenamefont {{J. Samuel}}(2000{\natexlab{a}})}]{Samuel-kritika1}%
  \BibitemOpen
  \bibfield  {author} {\bibinfo {author} {\bibnamefont {{J. Samuel}}},\
  }\href@noop {} {\bibfield  {journal} {\bibinfo  {journal} {Class. \ Quantum \
  Grav.}\ }\textbf {\bibinfo {volume} {17}},\ \bibinfo {pages} {L141} (\bibinfo
  {year} {2000}{\natexlab{a}})}\BibitemShut {NoStop}%
\bibitem [{\citenamefont {{J. Samuel}}(2000{\natexlab{b}})}]{Samuel-kritika2}%
  \BibitemOpen
  \bibfield  {author} {\bibinfo {author} {\bibnamefont {{J. Samuel}}},\
  }\href@noop {} {\bibfield  {journal} {\bibinfo  {journal} {Class. \ Quantum \
  Grav.}\ }\textbf {\bibinfo {volume} {17}},\ \bibinfo {pages} {4645} (\bibinfo
  {year} {2000}{\natexlab{b}})}\BibitemShut {NoStop}%
\bibitem [{\citenamefont {Livine}(2006)}]{Livine}%
  \BibitemOpen
  \bibfield  {author} {\bibinfo {author} {\bibfnamefont {E.~R.}\ \bibnamefont
  {Livine}},\ }\href@noop {} {} (\bibinfo {year} {2006}),\ \Eprint
  {http://arxiv.org/abs/gr-qc/0608135} {gr-qc/0608135} \BibitemShut {NoStop}%
\bibitem [{\citenamefont {{R. Geroch}}(1968)}]{Geroch}%
  \BibitemOpen
  \bibfield  {author} {\bibinfo {author} {\bibnamefont {{R. Geroch}}},\
  }\href@noop {} {\bibfield  {journal} {\bibinfo  {journal} {Jour. \ Math. \
  Phys.}\ }\textbf {\bibinfo {volume} {9}},\ \bibinfo {pages} {1739} (\bibinfo
  {year} {1968})}\BibitemShut {NoStop}%
\bibitem [{Note1()}]{Note1}%
  \BibitemOpen
  \bibinfo {note} {One may say that we can define the spinor structure locally
  and work with such structure. But there may occur some certain phatological
  features. We will not focus our attention to this problem. Therefore {"}no
  loss of generality{"}.}\BibitemShut {Stop}%
\bibitem [{\citenamefont {Fecko}(2004)}]{Fecko}%
  \BibitemOpen
  \bibfield  {author} {\bibinfo {author} {\bibfnamefont {M.}~\bibnamefont
  {Fecko}},\ }\href@noop {} {\emph {\bibinfo {title} {Diferenci\' alna
  geometria a Lieove grupy pre fyzikov}}}\ (\bibinfo  {publisher} {IRIS,
  Bratislava},\ \bibinfo {year} {2004})\ \bibinfo {note} {{[There exists
  english translation: M. Fecko - Differential Geometry and Lie Groups for
  Physicists (Cambridge University Press 2006)]}}\BibitemShut {NoStop}%
\bibitem [{Note2()}]{Note2}%
  \BibitemOpen
  \bibinfo {note} {One equation is still missing as we will see at the end of
  this section. But this equation is conservation of constraints given by (\ref
  {TorEQ}) and (\ref {EEQ}).}\BibitemShut {Stop}%
\bibitem [{Note3()}]{Note3}%
  \BibitemOpen
  \bibinfo {note} {$\protect \tmspace +\thinmuskip {.1667em}\protect \tmspace
  +\thinmuskip {.1667em}\protect \ensuremath {\protect \mathbf {e}}^a$ is
  coframe on $\protect \ensuremath {\protect \mathbb {T}}_1\protect \ensuremath
  {\protect \mathbf {M}}$, $\protect \mathaccentV {hat}05E{\protect \ensuremath
  {\protect \mathbf {e}}}^a$ is its representation on $\protect \ensuremath
  {\protect \mathbb {T}}_1\protect \ensuremath {\protect \mathscr
  {M}}$}\BibitemShut {NoStop}%
\bibitem [{Note4()}]{Note4}%
  \BibitemOpen
  \bibinfo {note} {Of course ADM formalism works with spatial metric $\protect
  \ensuremath {\protect \mathbf {q}}$ and therefore there are no coframe
  variables. For example in the Loop gravity Hamiltonian formulation starts
  with ADM, then orthonormal coframe $\protect \ensuremath {\protect \mathbf
  {e}}^i$ on $\Sigma $ is introduced and metric is expressed by orthonormality
  of this coframe, i.e. $\protect \ensuremath {\protect \mathbf {q}}$
  ($i,j=1,2,3$).}\BibitemShut {Stop}%
\bibitem [{\citenamefont {Wald}(1984)}]{Wald}%
  \BibitemOpen
  \bibfield  {author} {\bibinfo {author} {\bibfnamefont {R.~M.}\ \bibnamefont
  {Wald}},\ }\href@noop {} {\emph {\bibinfo {title} {General Relativity}}}\
  (\bibinfo  {publisher} {The University of Chicago Press},\ \bibinfo {year}
  {1984})\BibitemShut {NoStop}%
\bibitem [{\citenamefont {{P. A. M. Dirac}}(1950)}]{Dirac}%
  \BibitemOpen
  \bibfield  {author} {\bibinfo {author} {\bibnamefont {{P. A. M. Dirac}}},\
  }\href@noop {} {\bibfield  {journal} {\bibinfo  {journal} {Canad. \ J. \
  Math.}\ }\textbf {\bibinfo {volume} {2}},\ \bibinfo {pages} {129} (\bibinfo
  {year} {1950})}\BibitemShut {NoStop}%
\bibitem [{Note5()}]{Note5}%
  \BibitemOpen
  \bibinfo {note} {We omitted writing of details like $\forall \protect
  \mathaccentV {tilde}07E{\nu }^a\protect \dots $ in constraint's
  expressions.}\BibitemShut {Stop}%
\end{thebibliography}%

\end{document}